\documentclass[%
 reprint,
superscriptaddress,
 amsmath,amssymb,
 aps,
]{revtex4-2}

\usepackage{graphicx}
\usepackage{dcolumn}
\usepackage{bm}
\usepackage{siunitx}

\usepackage{booktabs}

\def\CuV{Cu\textsubscript{Zn}-V\textsubscript{S}}
\def\g2{g^{(2)}}

\begin{document}

\preprint{APS/123-QED}

\title{Room-temperature quantum emission from Cu\textsubscript{Zn}-V\textsubscript{S} defects in  ZnS:Cu colloidal nanocrystals}

\author{Yossef E. Panfil}
\affiliation{Department of Electrical and Systems Engineering, University of Pennsylvania, Philadelphia Pennsylvania 19104, USA}
\author{Sarah M. Thompson}
\altaffiliation{Present address: Sandia National Laboratories, Albuquerque, NM 87185}
\affiliation{Department of Electrical and Systems Engineering, University of Pennsylvania, Philadelphia Pennsylvania 19104, USA}
\author{Gary Chen}
\affiliation{Department of Chemistry, University of Pennsylvania, Philadelphia Pennsylvania 19104, USA}
\author{Jonah Ng}
\affiliation{Department of Electrical and Systems Engineering, University of Pennsylvania, Philadelphia Pennsylvania 19104, USA}
\author{Cherie R. Kagan}
\email[Corresponding authors. ]{lbassett@seas.upenn.edu \& kagan@seas.upenn.edu }
\affiliation{Department of Electrical and Systems Engineering, University of Pennsylvania, Philadelphia Pennsylvania 19104, USA}
\affiliation{Department of Materials Science and Engineering, University of Pennsylvania, Philadelphia Pennsylvania 19104, USA}
\affiliation{Department of Chemistry, University of Pennsylvania, Philadelphia Pennsylvania 19104, USA}
\author{Lee C. Bassett}
\email[Corresponding authors. ]{lbassett@seas.upenn.edu \& kagan@seas.upenn.edu }
\affiliation{Department of Electrical and Systems Engineering, University of Pennsylvania, Philadelphia Pennsylvania 19104, USA}

\date{\today}

\begin{abstract}
We report room-temperature observations of \CuV\ quantum emitters in individual ZnS:Cu nanocrystals (NCs).
Using time-gated imaging, we isolate the distinct, $\sim$3-$\mu$s-long, red photoluminescence (PL) emission of \CuV\ defects, enabling their precise identification and statistical characterization. 
The emitters exhibit distinct blinking and photon antibunching, consistent with individual NCs containing two to four \CuV\ defects.
The quantum emitters' PL spectra show a pronounced blue shift compared to NC dispersions, likely due to photochemical and charging effects.
Emission polarization measurements of quantum emitters are consistent with a $\sigma$-character optical dipole transition and the symmetry of the \CuV\ defect. 
These observations motivate further investigation of \CuV\ defects in ZnS NCs for use in quantum technologies.

\end{abstract}

\maketitle{}

\section{\label{sec:level1}Introduction}
Colloidal semiconductor nanocrystals (NCs) are prized for their size-tunable absorption and bright, stable luminescence \cite{Murray_Norris_bawendi,Sergent_science,Bruce_size,Guyot_ionnest_core_shell,Peng_epitaxial_growth_shell,Carbone_seeded_growth,Talapin_mix_dim,peng_100qy}. 
NCs are widely commercialized as biological tags \cite{alivisatos_bio_tagging} and as downconverters in displays \cite{Kagan_sargent, panfil_ACM,Panfil_Ang,Peng_displays,Sergent_science}.
They are being developed for use in optoelectronic LEDs, solar cells, and photodetectors \cite{Sargent_photovoltaic,Kagan_sargent,Tang2010,Sergent_science}.
NCs are also recognized for their excitonic  transitions as promising single-photon sources \cite{Nelson_SPS,Utzat_indistinguishable}.
For quantum technologies—such as quantum communication, quantum computing, and quantum sensing—colloidal NCs are advantageous platforms to introduce quantum point defects that feature long-lived quantum states that can be stored and manipulated \cite{Kagan2020ColloidalScience}.

Quantum point defects have emerged as key elements in quantum technologies \cite{Wolfowicz2021,Atature2018,Basset_kai_mei_2019}.
Certain point defects\,---\,typically formed by impurity atoms and their complexes with vacancies\,---\,introduce spin- and optically-active electronic states within the host material's band gap that can be manipulated and controlled.
Examples include the nitrogen-vacancy and silicon-vacancy centers in diamond, the divacancy in silicon carbide, and rare-earth impurities in metal oxides. 
Due to their isolation within suitably inert host materials, quantum defects can exhibit exceptionally long spin coherence times and coherent coupling to single photons \cite{Bar-Gill2013,Abobeih2018,Buckley_spin_light}. 
Quantum defects are typically created in bulk crystals using top-down processes, \textit{e.g.,} ion implantation and electron-beam irradiation. 
These processes are typically stochastic, and they are challenging to control at the nanoscale.
Moreover, they invariably cause unwanted damage that degrades the defects' quantum properties \cite{Ion_implantation}.

Colloidal NCs offer several advantages as hosts for quantum defects \cite{Kagan2020ColloidalScience}. 
They enable the bottom-up creation of defects \textit{via} wet chemical synthesis, facilitate defect localization within the typical sub-10 nm NC core, and allow exciton-wavefunction engineering by tailoring its size, shape, internal structure (\textit{i.e.}, creating core-shell and Janus architectures), and surface chemistry. 
olloidal NCs can be integrated into devices using solution-based assembly methods, enabling the precise placement of single NCs \cite{Kagan_Cossairt_arrays,Guymon_print,Saboktakin2013,Henry_template} and the organization of ordered NC arrays \cite{ordered_3d_superlattice}.

A few previous studies have explored the introduction of impurities in NC host materials, such as Mn$^{2+}$ in CdTe NCs \cite{Mn_5_2} and Er$^{3+}$ in CeO\textsubscript{2} NCs \cite{Er_Ceo2}.
Recently, we reported the synthesis and optical characterization of copper-doped ZnS (ZnS:Cu) NCs, which feature red emission due to \CuV\ defects, \textit{i.e.}, where a copper atom replaces a zinc atom adjacent to a sulfur vacancy \cite{Thompson_red_emission}. 
Here, we isolate individual ZnS:Cu NCs and use time-gated imaging to identify quantum emission from \CuV\ defects within the NCs.
Despite the long, $\sim$3 $\mu$s emission lifetime that limits their brightness, we observe blinking and photon anti-bunching behavior consistent with NCs that contain just two to four defects.
We find that the \CuV\ quantum emitters luminesce exclusively at red wavelengths, without the blue emission tail observed from ensembles.
Interestingly, the red quantum emission is blue-shifted relative to that seen in ensemble spectra, and time-dependent studies suggest that photochemistry and charging play an important role.
The isolation of individual NCs further enables a statistical study of emission polarization from the quantum emitters, which indicates a $\sigma$-dipole character of the optical transitions and supports the energy-level structure proposed for \CuV\ defects based on symmetry analysis and previous calculations \cite{Thompson_red_emission, Shionoya1966NatureMeasurements}. 
Our findings advance the development of quantum defects within colloidal NCs, laying a foundation for their future integration into quantum devices and applications.

\begin{figure*}
\includegraphics[scale=1]{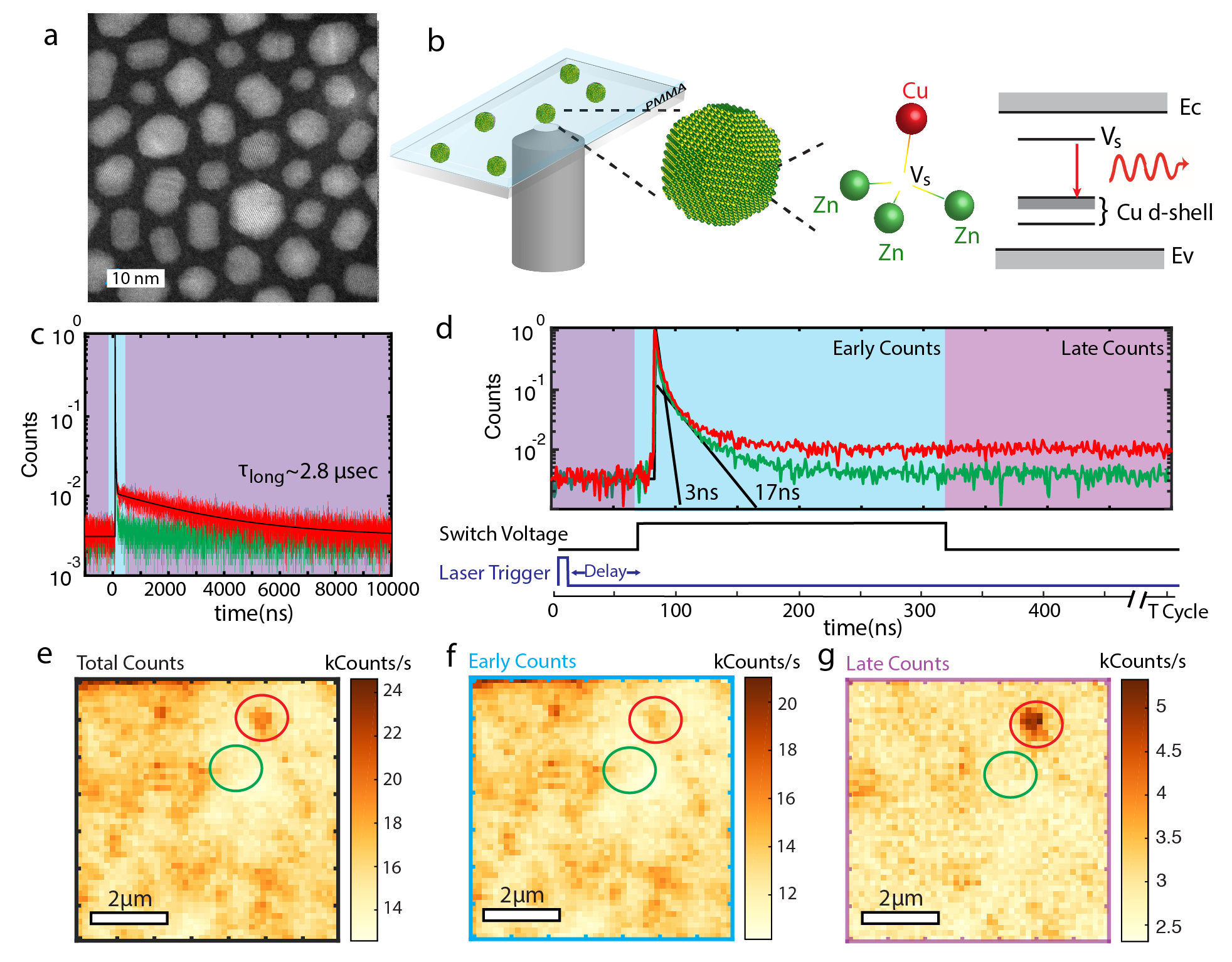}
\caption{\label{fig:Synthesis and Structure}
 \textbf{Structural and optical characterization of individual ZnS:Cu NCs} 
 (a)  High-resolution scanning transmission electron micrograph of ZnS:Cu NCs.
 (b) Schematic of the confocal microscope used to probe individual colloidal ZnS:Cu NCs and the \CuV\ defect structure along with its energy levels. 
 (c) Emission decay curves after excitation by a 405-nm laser pulse (red and green data) obtained from two positions on a sample, indicated by red and green circles in (e-g), respectively.
 (d) Zoomed-in view of the decay curve from (c) immediately following excitation, along with the time sequence used to separate late photon counts (purple shaded region) from early photon counts (blue shaded region).
 (e-g) A representative confocal PL image is resolved from total counts (e) into separate images for only early (f) and late (g) counts.
}
\end{figure*}

\section{Results and Discussion}

\subsection{Synthesis of red-emitting ZnS:Cu NCs}
Colloidal Cu-doped ZnS NCs are synthesized following a previously-described approach to produce red emission characteristic of \CuV\ defects, also known as R-Cu emission \cite{Thompson_red_emission}.
Briefly, the ZnS NCs are synthesized using the single-source precursor approach developed by Zhang \textit{et al.} \cite{Zhang2010}, where zinc diethyldithiocarbamate (Zn(Ddtc)\textsubscript{2}) is thermally decomposed in oleic acid (OA) and oleylamine (OM).
The ZnS NCs are doped with Cu during synthesis by adding Cu(CH\textsubscript{3}COO)\textsubscript{2}$\cdot$H\textsubscript{2}O dissolved in ultra-pure DI water, at a concentration corresponding to a Cu molar ratio of 0.1\%, into the synthesis pot. 
The ZnS:Cu NCs are isolated from the growth medium and dispersed in hexane to prepare 10 mg/mL ZnS:Cu NC dispersions.
High-resolution scanning transmission electron microscope (STEM) images of ZnS:Cu NCs show that the samples consist predominantly of crystalline hexagonal plates with diameters of 6.60$\pm$1.43 nm and heights of 5.10$\pm$0.67 nm (Fig. \ref{fig:Synthesis and Structure}a). 
The observed hexagonal geometry and X-ray diffraction (XRD) measurements are consistent with NCs having a wurtzite-10H crystal structure (Supporting Information Fig. S1) \cite{Zhang}.

\subsection{Single NC isolation}

Colloidal ZnS:Cu NC dispersions are diluted in a 3 wt\% solution of PMMA (Poly(methyl methacrylate)) in toluene and deposited by spin-coating onto glass coverslips for confocal imaging and spectroscopic measurements (Fig. \ref{fig:Synthesis and Structure}b).
The use of PMMA is seen to increase the PL stability enabling long (\textgreater 1 h) measurements and analysis \cite{PMMA_blueing}.
In order to isolate single NCs, the NC dispersion is diluted three times at 1:10 NC dispersion:PMMA solution. After each dilution step, scanning confocal microscopy images are acquired using a 405 nm excitation laser.
The data indicate a decrease in the background intensity and concentration of \CuV\ defects (resolved using time-gated imaging, as described in the next section) with each successive dilution.
Figures~\ref{fig:Synthesis and Structure}e-g correspond to a single dilution; see the Supporting Information Fig. S2 for representative images of other dilutions. 
The stepwise dilution is crucial for reducing the NC concentration to a level where individual quantum emitters can be resolved while confirming that the emitters are associated with the ZnS:Cu NC sample and are not artifacts of the substrate. 

\subsection{Time-gated confocal microscopy}

The \CuV\ defect has an emission lifetime of several microseconds at room temperature \cite{Thompson_red_emission}.
This relatively long lifetime limits the emitters' brightness, but it also distinguishes them from background sources of PL, which have much shorter (typically $<$20 ns) lifetimes. 
Figure \ref{fig:Synthesis and Structure}c presents PL emission decay data acquired from two positions on the sample, marked in Figure \ref{fig:Synthesis and Structure}e by red and green circles. 
The measurements track emission at wavelengths $\geq$ 450 nm following a 405-nm laser pulse.
At the green-marked position, we only observe two short-lifetime components of 3 ns and 17 ns (Fig. \ref{fig:Synthesis and Structure}d), consistent with sub-bandgap emission from vacancies and interstitials in ZnS \cite{Thompson_red_emission} and substrate emission.
At the red-marked position, in contrast, we observe an additional 2.8 $\mu$s lifetime component, consistent with \CuV\ emission \cite{Thompson_red_emission,Yen2007}.

We apply a time-resolved detection method (Fig. \ref{fig:Synthesis and Structure}d) that separates photon counts into early (blue; detected within $\sim$200 ns of the laser pulse) and late (purple; detected more than $\sim$200 ns after the laser pulse) windows; see Methods for additional details.
The early-counts image (Fig. \ref{fig:Synthesis and Structure}f) predominantly shows short-lifetime emission from the substrate or other defects, whereas the late-counts image (Fig. \ref{fig:Synthesis and Structure}g) distinguishes the \CuV\ defect emission from within the ZnS:Cu NCs.
This approach overcomes one of the main challenges in studying transition-metal-based emitters, which often feature long emission lifetimes \cite{Cr4+}. 
As shown in the following sections, this method further enables detailed investigations of the temporal and spatial emission characteristics of individual emitters.




\begin{figure}
\includegraphics[scale=0.4]{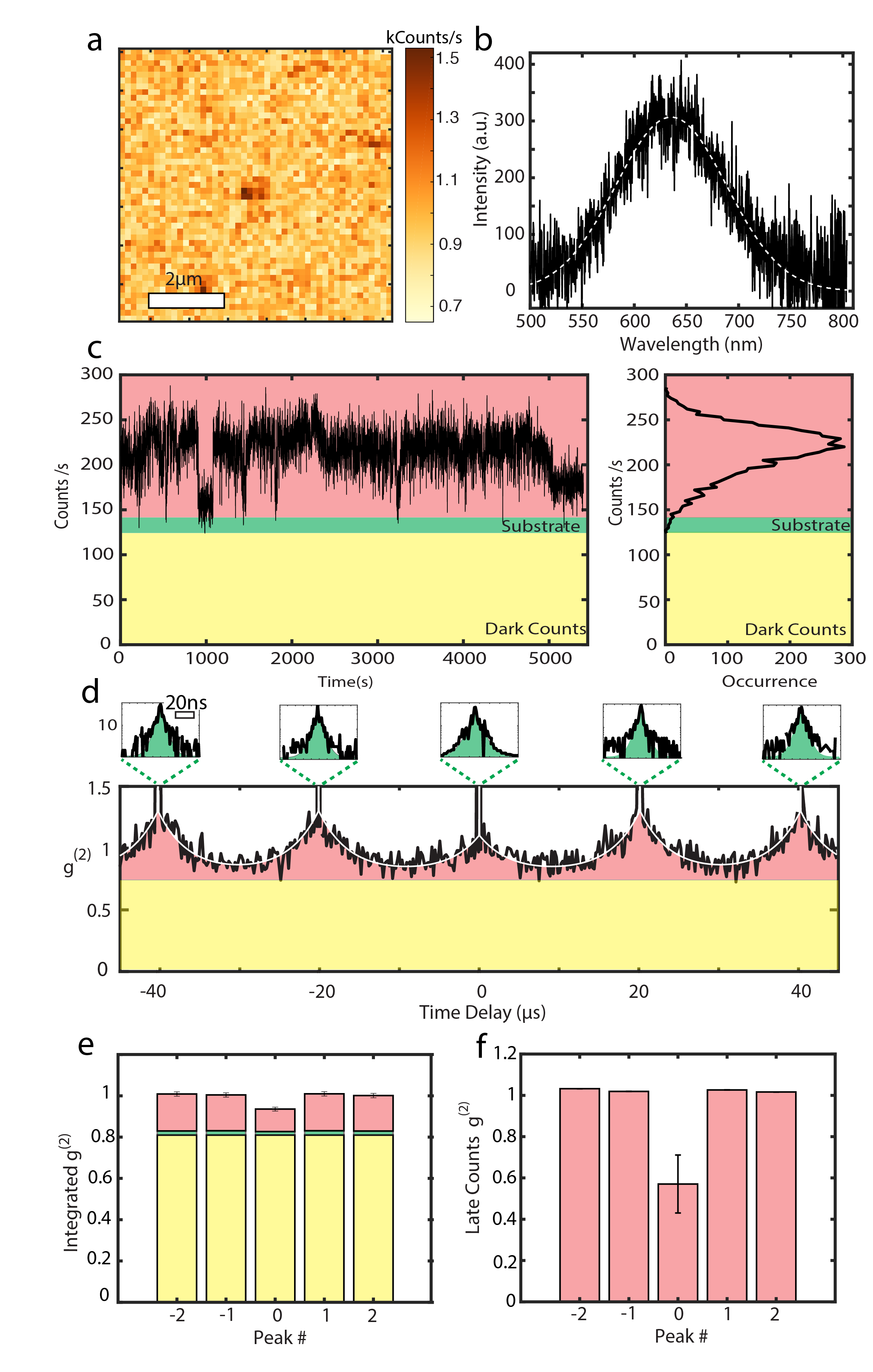}
\caption{\label{fig:Quantum Emission from Cu-Vs defects in ZnS:Cu NCs}
 \textbf{Quantum emission.} 
 (a) Late-counts image of a ZnS:Cu NC sample dispersed on a glass coverslip, prepared after three dilutions at 1:10 NC dispersion:PMMA solution. 
 From the dim spot in the center of (a), we collect the
 (b) PL spectrum, shown also with a Gaussian fit (dashed curve) as described in the text;
 (c) time trace of PL emission intensity (left) and intensity histogram (right) during 5400 s of a  $\g2$ measurement, in which the substrate emission and  dark-count intensity levels are marked in green and yellow, respectively;
 and (d) the $\g2$ function, fit using a periodic biexponential decay model (white curve).
 Small panels in (d) are zoomed-in views of the short-timescale $\g2$ function at each peak position.
 (e) Integrated, discretized $\g2$ function representing integrated counts under each peak in (d), resolved to show the long component (red), the short component (green), and the offset/dark counts (yellow).
(f) $\g2$ function for the long-lifetime component only, corrected for background effects as described in the text and Supporting information.
}
\end{figure}

\subsection{Observation of quantum emission}

Figure \ref{fig:Quantum Emission from Cu-Vs defects in ZnS:Cu NCs}a presents a representative late-counts image 
 of a sample prepared after three dilution steps.
These samples show dim, isolated spots over a diffuse background that is dominated by detector dark counts.
The spectrum collected from one such spot is shown in Figure \ref{fig:Quantum Emission from Cu-Vs defects in ZnS:Cu NCs}b, along with a fit using a Gaussian peak centered at 635 nm with a full-width at half maximum (FWHM) of 125 nm.
This emission profile is consistent with the red emission from \CuV\ defects in ZnS \cite{Thompson_red_emission}.
The PL spectra are not time gated, so they include both early and late photons. 
Nonetheless, the spectra collected from dim spots lack the blue emission tail observed from ensembles of ZnS NCs (both Cu-doped and undoped).
This observation suggests that the blue PL in ensemble measurements originates from particles that do not contain  \CuV\ defects and is consistent with samples having a distribution of both doped and undoped particles. 
The blue emission is likely suppressed in doped particles since excitations from those higher-energy states non-radiatively relax before decaying through the red \CuV\ optical transition, which dominates the spectrum. 


Figure \ref{fig:Quantum Emission from Cu-Vs defects in ZnS:Cu NCs}c shows the PL emission intensity from the dim spot in Figure~\ref{fig:Quantum Emission from Cu-Vs defects in ZnS:Cu NCs}a over 90 min (5400 s) in response to 405~nm laser pulses applied at a 50 kHz repetition rate.
The PL is remarkably stable over this collection window, and it exhibits stochastic intensity changes known as blinking, where the intensity briefly drops to lower intensity levels, including to background levels that are dominated by detector dark counts (yellow level in Fig.~\ref{fig:Quantum Emission from Cu-Vs defects in ZnS:Cu NCs}c), before returning again to a brighter steady state.
Blinking is a hallmark of quantum emitters \cite{Nirmal1996}.
Notably, the PL intensity trace in Figure \ref{fig:Quantum Emission from Cu-Vs defects in ZnS:Cu NCs}c is neither spectrally nor temporally filtered.
For these excitation settings, apparently, the PL signal above detector dark counts is dominated by emission from a small number of \CuV\ defects acting as quantum emitters.


In order to quantify the number of emitters, we use a Hanbury-Brown-Twiss interferometer to study the photon-emission autocorrelation function, $\g2$.
Whereas continuous-wave excitation is typically used to confirm quantum emission and study the optical dynamics of bright emitters with short lifetimes \cite{Fishman_PRXQuantum}, here we employ pulsed excitation in order to isolate the long-timescale signals from \CuV\ defects from the short-timescale background emission as well as from the uncorrelated dark counts.
Figure \ref{fig:Quantum Emission from Cu-Vs defects in ZnS:Cu NCs}d shows a representative $\g2$ measurement from 5400 s of data collection, where the peaks separated by 20.13 $\mu$s correspond to the laser repetition period.
The data are fit using a periodic function consisting of symmetric, biexponential peaks\,---\,with best-fit decay times of 5$\pm$1 ns and 3.5$\pm$0.11 $\mu$s, respectively\,---\,together with a constant background due to detector dark counts.
While the decay times are fit uniformly across each peak, the amplitudes corresponding to each component in the model are allowed to vary.
We find that the short component (green; see insets in Fig.~\ref{fig:Quantum Emission from Cu-Vs defects in ZnS:Cu NCs}d) exhibits a constant amplitude independent of delay, whereas the long component exhibits a marked reduction in amplitude for the peak centered at zero delay, compared to the satellite peaks at non-zero delay.
The reduced likelihood to detect two long-lived photons from the same excitation pulse rather than from two separate pulses is known as photon antibunching, and it directly indicates the quantum nature of the \CuV\ defect emission in these samples. 


A quantitative analysis of the $\g2$ measurements requires an accounting of each of these temporally-resolved contributions, together with interactions between. For example, an event in the $\g2$ data may represent a situation where one photon is emitted by the sample or substrate, and the other is a detector dark count.
Figure \ref{fig:Quantum Emission from Cu-Vs defects in ZnS:Cu NCs}e displays the integrated, discretized $\g2$ value associated with each laser pulse, resolved according to lifetime.
The integrated contribution from the short component (green) in the side peaks is 11\% of the long-lifetime contribution (red), consistent with their respective PL intensities in Figure \ref{fig:Quantum Emission from Cu-Vs defects in ZnS:Cu NCs}c.
As such, substrate emission has a negligible effect on the discretized $\g2$.
The contribution of dark counts (yellow) is more significant.
The uncorrelated background in Figure~\ref{fig:Quantum Emission from Cu-Vs defects in ZnS:Cu NCs}d represents situations where either one or both photons comprising the event are dark counts, and these processes have similar likelihood since the dark count rate is comparable to the overall emission rate from \CuV\ defects in this experiment. 

Figure \ref{fig:Quantum Emission from Cu-Vs defects in ZnS:Cu NCs}f shows the discretized $\g2$ function for the long-lifetime component only, corrected for the contributions from background and dark counts; see Methods and Supporting Information for details on the correction.
We obtain a corrected value from the central peak of $\g2(0)=0.58 \pm 0.14$.
For $N$ emitters of equal intensity, we expect $\g2(0)=1-1/N$, hence this measurement is consistent with the presence of two or three emitters of similar intensity in this particular spot.
The Supporting Information includes other examples of photon antibunching, with $\g2(0)$ values ranging from 0.52 to 0.8, consistent with spots containing two, three, or four \CuV\ defects.

\begin{figure}
\includegraphics[width=0.5\textwidth]{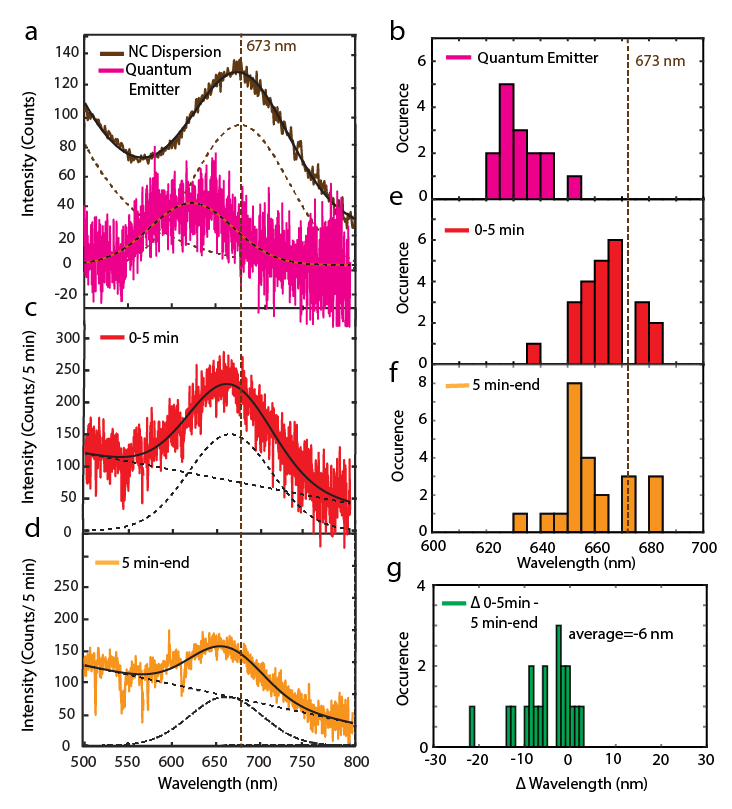}
\caption{\label{fig:PL Spectrums}
\textbf{PL spectra and response to illumination}
(a) PL spectrum of a quantum emitter (magenta data) acquired for 5 min together with that of a NC dispersion (brown data) acquired pointwise for 2.5 sec per wavelength. Dashed curves represent Gaussian fits to the red spectral components and a decaying exponential fit to the blue tail of the NC dispersion data. 
(b) Histogram of the best-fit central wavelength from spectra of 15 suspected quantum emitters, some of which have $\g2(0)$ values less than 1 and others with the same emission intensity. 
(c,d) PL spectra collected from a bright spot during the first 5 min of illumination (c) and during the subsequent 35 min of illumination (d). Dashed curves represent Gaussian fits to the red spectrum and a linear function to capture the blue tail.
(e,f) Histograms of the central emission peaks for 24 bright spots during the first 5 min (e) and after 5 min of illumination (f).
(g) Difference in the central PL wavelength for spectra collected in the first 5 min of illumination compared to those collected subsequently.
}
\end{figure}

\subsection{Spectral properties and stability}

The isolation of single ZnS:Cu NCs containing just a few quantum emitters enables the statistical characterization of individual \CuV\ defects and comparisons with ensemble experiments.
Figure \ref{fig:PL Spectrums}a overlays the spectrum of a quantum emitter (with $\g2(0)<1$) with that of a NC dispersion. 
The quantum emitter spectrum, fit using a Gaussian distribution centered at 623 nm with a FWHM of 112 nm, is notably blue-shifted compared to the dispersion spectrum, which peaks at 673 nm with a FWHM of 127 nm.
Measurements of 15 quantum emitters (Fig. \ref{fig:PL Spectrums}b) reveal central wavelengths ranging from  620–655 nm, which are consistently blue shifted relative to the NC dispersion spectrum.

We attribute this blue shift to the higher laser intensity and longer exposure times needed to acquire single-NC spectra, which likely induce photochemical changes or charging effects that modify the emission spectrum \cite{photochemistry_2018,Klimov_blue_shift,Peng_blue_shift,Nozik,Nair}.
This hypothesis is supported by illumination-dependent measurements of bright spots identified through fast PL scans across large areas.
These bright spots exhibit $\g2(0)=1$ and likely include many ZnS:Cu NCs.  
Figures \ref{fig:PL Spectrums}c,d show representative PL spectra from such a bright spot collected during the first 5 min of illumination and during a subsequent collection window from 5 -- 40 min.
Both spectra can be decomposed into a linear component consistent with the blue-tail emission from undoped NCs and a Gaussian peak attributed to emission from \CuV\ defects. 
The PL intensity of the red part of the spectrum decreases significantly during the acquisition. 
Moreover, the peak of the red emission blue-shifts; the center wavelength moves from 666$\pm$2.4 nm to 660$\pm$2.4 nm.

Histograms of the best-fit emission peaks for 24 bright spots measured under identical procedures (Figs.~\ref{fig:PL Spectrums}e,f) show a distribution of peak wavelengths initially between 650--680 nm that subsequently shift to shorter wavelengths.
The shifts, shown in Figure~\ref{fig:PL Spectrums}g, range from -20 nm to +3 nm, with a mean of -6 nm.
The FWHM for the red component typically falls between 80--160 nm, and, notably, it does not systematically change following illumination; see Fig.~S6 in the Supporting Information. 


Previous studies of ZnS in bulk and NC forms reported Cu-related red emission center wavelengths ranging between 600 nm and 700 nm \cite{Shionoya1964NatureCrystals,640nm_1967, Bowers_melamed, Aven_1958, Bol2002,Thompson_red_emission}.
The wide variation suggests the existence of multiple classes of \CuV\ defects, \textit{e.g.}, with different charge states or local environments.
In this context, the observed blue shift over time can be interpreted in two ways: (1) a predominantly red-emitting population may bleach under illumination, leaving behind a more stable population with a blue-shifted spectrum, or (2) the red population may undergo a transformation into a blue-shifted state. 
In either case, the data suggest that the blue-shifted \CuV\ defect emission represents a more stable configuration under illumination. 
The fact that the FWHM does not change seems to support the second option of an overall shift.
Moreover, the FWHM of quantum emitters (typically 95--145 nm; see Fig.~S6 in the Supporting Information) is consistent with that of the bright spots and the dispersion spectra.
This implies that the broad ensemble linewidth arises primarily from homogeneous broadening.


\begin{figure}
\centering
\includegraphics[scale=1.3]{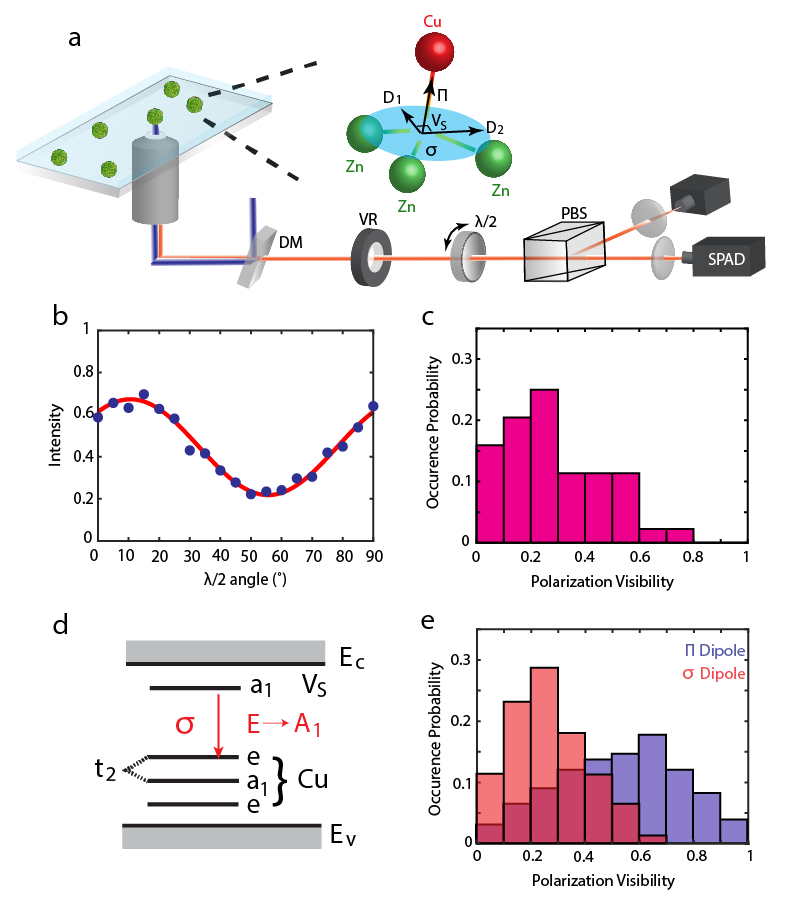}
\centering
\caption{\label{fig:Cu-Vs Defects Red Emission Dipole Structure}
\textbf{Emission polarization measurements} 
(a) Optical setup used for emission polarization measurements (DM: Dichroic mirror; VR: variable retarder; $\lambda/2$: half-wave plate; PBS: polarizing beamsplitter; SPAD: single-photon avalanche detectors).
The inset shows the \CuV\ defect along with possible optical transition dipoles: a $\pi$ dipole polarized along the axis connecting the V\textsubscript{S} and Cu atom, and a $\sigma$ dipole consisting of two orthogonal transitions $D_1$ and $D_2$ polarized perpendicular to the defect axis.
(b) Late-counts, differential emission intensity as a function of the $\lambda$/2 waveplate angle; the polarization visibility for this spot is  $P=0.51\pm0.11$.
(c) Distribution of emission polarization visibility extracted from measurements of 44 dim spots.
(d) Electronic level structure of the \CuV\ defect in a $C_{3v}$-symmetric configuration.
(e) Simulated emission polarization visibility distributions for randomly oriented pairs of emitters characterized by a $\pi$-dipole transition (purple) and a $\sigma$-dipole transition (red).
} 
\end{figure}

\subsection{Emission polarization and implications for electronic structure}

The emission and absorption polarization of quantum defects reveal important information regarding their electronic structure and optical dipole transitions; see Fig.~4. 
In particular, polarization visibility can distinguish between optical transitions consisting of a single, linearly-polarized optical dipole (known as a $\pi$ dipole), and those featuring two degenerate or closely spaced orthogonal transition dipoles (known as a $\sigma$ dipole).
For the \CuV\ defect, the $\pi$ and $\sigma$ dipoles correspond to transitions polarized along or perpendicular to the axis connecting the copper and vacancy positions, respectively (see Fig.~\ref{fig:Cu-Vs Defects Red Emission Dipole Structure}a).
These polarization patterns are hidden in measurements of large ensembles, but they can be resolved through measurements of individual particles, even if they are randomly oriented \cite{ZnO_defect}.

Here we study emission polarization, since the excitation occurs through higher-lying energy levels close to the band edge whose transition dipoles are decoupled from the defect states.
Using pulsed excitation at 405 nm with circular polarization, we apply the time-gating method to isolate the long-lifetime emission from \CuV\ defects and analyze its polarization. 
As shown in Figure~\ref{fig:Cu-Vs Defects Red Emission Dipole Structure}a, we employ a $\lambda$/2 waveplate and polarization beam splitter (PBS) to determine the linear polarization visibility.
The use of a PBS and two detectors makes the polarization a differential measurement, thus mitigating noise associated with the blinking behavior of quantum emitters; see Sec.~VI and Fig.~S7 in the Supporting Information for details.

Figure \ref{fig:Cu-Vs Defects Red Emission Dipole Structure}b presents emission polarization data for a representative quantum emitter as a function of the $\lambda$/2 waveplate angle, together with a sinusoidal fit. 
From the fit, we determine the polarization visibility, \begin{equation} \label{eq:polarization visibility}  
P=\frac{I_\mathrm{max}-I_\mathrm{min}}{I_\mathrm{max}+I_\mathrm{min}},
\end{equation}
in terms of the maximum ($I_\mathrm{max}$) and minimum ($I_\mathrm{min}$) intensity. 
Figure~\ref{fig:Cu-Vs Defects Red Emission Dipole Structure}c shows the distribution of $P$ from measurements of 44 dim spots, all of which exhibit PL intensity similar to confirmed quantum emitters with $\g2(0)$ values below 1. 
We expect each of these spots to include two to four \CuV\ defects.

To interpret these measurements, Figure~\ref{fig:Cu-Vs Defects Red Emission Dipole Structure}e shows the expected polarization visibility distribution for randomly oriented pairs of emitters that feature either $\pi$ dipoles or $\sigma$ dipoles.
The simulations that produced Fig.~\ref{fig:Cu-Vs Defects Red Emission Dipole Structure}e are described in Sec.~VII of the Supporting Information.
They account for the 1.3 numerical aperture of the collection optics, and we also include simulations for individual emitters, as well as for $\sigma$ dipoles with unequal contributions.
The simulations consistently show that $\pi$ dipoles produce distributions with higher average visibility, and with a range of visibility extending close to unity.
In contrast, $\sigma$ dipoles (whether equal or unequal) produce distributions with lower visibility and values primarily with $P<0.5$.
The experimental data align closely with the theoretical expectations for $\sigma$ dipoles.

Figure~\ref{fig:Cu-Vs Defects Red Emission Dipole Structure}d shows the prevailing molecular orbital theory model of the \CuV\ defect \cite{Shionoya1966NatureMeasurements,Thompson_red_emission}.
A set of low-lying orbitals mainly arising from the Cu $d$-shell are split by the crystal field, and a higher-lying orbital level arises from the S vacancy. 
For defects with $C_{3v}$ symmetry, which characterizes all \CuV\ defects along the body diagonal (111) in zinc-blende crystals and those aligned along the $c$-axis (0001) in wurtzite crystals, the Cu $d$ levels are split into $e$-, $a_1$-, and $e$-symmetry orbitals as shown.

The nature of the optical ground and excited states depends on the charge state of the defect.
The predominant charge state of the \CuV\ defect is unknown, and the optical cycle likely involves multiple charge states \cite{Thompson_red_emission}.
Nonetheless, whether the defect is neutral, positive, or negative with respect to the lattice, the Cu ion is expected to have the $3d^{10}$ configuration in its optical ground state.
In the optical excited state, a hole is created in the uppermost Cu level(s), and the characteristic red luminescence results from recombination with an electron in the $V_\mathrm{S}$ level \cite{Shionoya1966NatureMeasurements,Thompson_red_emission}.
Hence, the ground state has $A_1$ symmetry whereas the optical excited state has $E$ symmetry. 
Optical emission from $E\rightarrow A_1$ occurs via a $\sigma$ dipole, in agreement with our measurements.

\subsection{The role of crystal structure}

As discussed previously, STEM and XRD structural characterization measurements imply that our samples are predominantly wurtzite, likely with the 10H crystal symmetry previously reported \cite{Zhang}. 
In wurtzite crystals, the \CuV\ defect can exist in multiple lattice configurations with distinct optical properties.
The possible configurations are analogous to well-known point defects like the divacancy and nitrogen-vacancy center in SiC \cite{Gordon_2015,Davidsson_2018,Miao_2019}.
This inhomogeneity may influence the distribution of spectral emission peaks observed in Figure~\ref{fig:PL Spectrums}. 
At this stage, however, it is not possible to distinguish the effects of distinct defect configurations from potential photophysical mechanisms that can also shift the emission energy.
Furthermore, orange-red emission spectra from ZnS:Cu bulk crystals show negligible differences when comparing 
zinc-blende \textit{vs.} wurtzite crystals \cite{Aven_1958}. 
This insensitivity to crystal structure might imply preferential formation of certain defect configurations in hexagonal polytypes, that the crystal field perturbations associated with different configurations are similar, or it may result from the overall broad emission band, which could obscure variations due to different defect configurations. 

Crystal symmetry also has potential implications for the defect's electronic structure and polarization patterns. 
In wurtzite crystals, $c$-axis defect configurations with $C_{3v}$ symmetry are described by Figure~\ref{fig:Cu-Vs Defects Red Emission Dipole Structure}d, whereas basal configurations have lower ($C_{1h}$) symmetry, and the degeneracy of the $e$ states is further lifted.
In the basal case, the optical excited state splits into two levels which respectively connect to the ground state through orthogonal optical dipoles. 
If the splitting is large and the emission is dominated by only one transition, a $\pi$-dipole pattern is expected. 
We expect, however, that the symmetry breaking of basal-plane defects can be treated as a relatively weak perturbation to the $C_{3v}$ states; this is the case for analogous divacancy defects in wurtzite SiC \cite{Miao_2019}, and the $d$-orbital wavefunctions of \CuV, being more localized than for a divacancy, should be even less sensitive to the crystal field. 
In this case, both transitions contribute to the emission, and we still expect a $\sigma$-dipole pattern; see Sec.~VIII of the Supporting Information for simulations of this situation.
In conclusion, our observations of $\sigma$-polarized emission patterns generally support the interpretation of red emission in ZnS:Cu NCs as arising from \CuV\ defects with predominantly $C_{3v}$ symmetry.

\section*{Conclusion}

In this study, we report the potential of \CuV\ defects in ZnS:Cu NCs as robust quantum emitters and study their optical properties and electronic structure using single-NC spectroscopy.

Looking ahead, these results inform exciting prospects for optical and spin control using \CuV\ defects in ZnS:Cu NCs. 
Certain charge configurations are predicted to be paramagnetic, with a highly localized spin that can be optically or magnetically controlled similar to other transition-metal-based quantum systems \cite{Cr4+,Vanadium,Ni_MgO}.
To harness this potential, further investigation into the spin and optical dynamics of \CuV\ defects is required. 
Given the large difference between the excitation and emission wavelengths used in this study, the optical cycle likely involves ionization and recombination. 
Alternative schemes, particularly resonant excitation at low temperatures, may be important for achieving spin initialization and readout. 
These efforts will pave the way for understanding and exploiting the defect’s spin properties for spin-photon interfaces and quantum sensing.

Beyond \CuV\ defects, this work sets the stage for the rational design of other transition-metal-vacancy complexes in ZnS, leveraging the versatility of colloidal NC synthesis. 
The choice of transition metal determines both the optical and spin properties of the defect through the occupation of the $d$-shell.
While ZnS stands out as an excellent quantum host material due to its wide bandgap, low nuclear spin density, and weak spin-orbit interaction, it may also be interesting to consider other NC materials \cite{Kagan2020ColloidalScience}.
 
The use of colloidal NCs as host materials further expands the scope of quantum-defect research. 
NCs can be co-assembled, printed or assembled in arrays on substrates, and integrated into photonic and plasmonic structures \cite{Cui2019,Kagan_Cossairt_arrays,Guymon_print,Saboktakin2013,Henry_template,ordered_3d_superlattice}. 
These advantages, combined with the facile, scalable synthesis and tunability of NCs, position them as a unique platform for advancing nanophotonics and quantum engineering.


\section{Methods}

\subsubsection{Synthesis of colloidal ZnS:Cu NCs}

Zinc diethyldithiocarbamate (Zn(Ddtc)\textsubscript{2}), copper (II) acetate monohydrate (Cu(CH\textsubscript{3}COO)\textsubscript{2}$\cdot$H\textsubscript{2}O), oleic acid (OA, 90\% purity), oleylamine (OM, 70\% purity) are purchased from Sigma-Aldrich. All chemicals are used without further purification. A 10 mL solution of 5 mM Cu(CH\textsubscript{3}COO)\textsubscript{2}$\cdot$H\textsubscript{2}O dissolved in deionized water is prepared. 0.1 mL of this solution is then added to a 100 mL three-neck flask containing 40 mmol OM, 40 mmol OA, and 0.4 mmol Zn(Ddtc)\textsubscript{2}. The mixture is heated at 120  $^{\circ}$C under vacuum for 60 min. The vessel is then heated to 300 $^{\circ}$C under a nitrogen atmosphere and maintained at 300 $^{\circ}$C for 45 min to nucleate and grow the NCs. 
The reaction pot is removed from heat and left to cool to 60 $^{\circ}$C. 
The NCs are collected \emph{via} addition of ethanol and centrifugation for 3 cycles before being re-dispersed in hexane to a concentration of 10 mg/mL. 

\subsubsection{Sample preparation}
The ZnS:Cu NC dispersion (10 mg/mL) is diluted in a 3 wt\% PMMA in toluene solution.
The diluted dispersion is then spin-coated onto a cleaned glass coverslip at 4,000 rpm for 60 s.
The coverslip is subsequently mounted on an inverted microscope for optical characterization.

\subsubsection{Confocal microscopy}
The PL properties of the ZnS:Cu NCs are studied using a custom-built time-gated confocal microscopy setup.
A pulsed 405 nm laser (pulse duration: 100 ps, repetition rate: 10 kHz) is used to excite the sample.
The excitation and emission light are focused and collected using a 100x oil immersion objective (NA = 1.4).
The emission is separated from the excitation light using a dichroic mirror and directed to one or more single-photon avalanche photodiodes (SPADs) using a dichroic mirror.
Confocal images of the dilute ZnS:Cu NC assemblies are acquired by scanning the optical excitation and collection spot using a fast steering mirror.
For the images in Fig.~\ref{fig:Synthesis and Structure}, the scan area is 7 $\mu$m $\times$ 7 $\mu$m, the pixel resolution is 150~nm, and the dwell time is 2 ms per pixel.

\subsubsection{PL lifetime measurements}
For lifetime measurements, photon detection signals from the SPAD are directed to a Time-Correlated Single Photon Counting (TCSPC) system.
The TCSPC system records a histogram of the time difference between excitation pulses and photon detection.

\subsubsection{Time-gated imaging}
Time-gated imaging is implemented by routing the electronic photon detection signals through a series of rf switches and then into a set of counters.
An arbitrary waveform generator (AWG), controls both the excitation laser pulses and the switching of the detection system.
As described in the main text, photon detection events are separated into early ($<$260 ns after excitation) and late counts ($>$260 ns after excitation). 
In this way, early- and late-count signals can be resolved as separate images when scanning the confocal microscope.

\subsubsection{Autocorrelation measurements}
In autocorrelation measurements, collected photons are split by a 50:50 fiber beamsplitter into two SPADs, and the photon detection signals are directed to a TCSPC system operating in time-tagged mode, such that the arrival time of each photon is recorded.
The autocorrelation function is subsequently calculated from the full dataset, along with the averaged emission intensity as a function of time \cite{Fishman_PRXQuantum}. 

The autocorrelation data are fit using a sum of symmetrically decaying exponential functions, with two lifetime components for each laser pulse, together with a constant background to capture dark-count contributions.
The long and short lifetimes are common fit parameters for each pulse, while the corresponding amplitudes are free to vary.
Subsequently, the best-fit $\g2(\tau)$ curve is converted to a discrete, lifetime-resolved $\g2$ function by integrating the contributions from dark counts, short-lifetime events, and long-lifetime events.
In order to extract the effective $\g2$ function corresponding only to the long-lifetime process, a correction is applied to account for events where one photon is a dark or short-lifetime count.
The relationship between the measured  $\g2$ value, $g_{mL}^{(2)}$, and the corrected value, $g_{L}^{(2)}$, is:
\begin{equation} \label{g2 correction} 
g_{L}^{(2)} = [g_{mL}^{(2)} (S_{L}+S_{S}+D)^{} - S_{S}S_{L} ] / S_{L}^2
\end{equation} 
where $S_L$ is the time-averaged intensity of the long component$S_S$ is the intensity of the short component, and $D$ is the time-independent background (see supporting information). 

\subsubsection{Polarization measurements}
For emission polarization measurements, the collected photons are directed through a $\lambda/2$ plate and PBS into two SPADs; see Fig.~\ref{fig:Cu-Vs Defects Red Emission Dipole Structure}. 
The 405~nm excitation laser is prepared with circular polarization, and the time-gating method is used to  isolate the long-lifetime red emission from the \CuV\ defects.
The intrinsic birefringence of the optical setup is compensated using a liquid crystal variable wave plate in the collection path, whose retardance is adjusted to preserve the linear polarization of a calibration source tuned to 650~nm.
With this correction, polarization visibility curves are acquired by rotating the angle of the $\lambda$/2 waveplate and recording the time-gated signal from the two SPADs.
Compared to the typical approach of measuring only a single polarization component (\textit{e.g.,} using a linear polarizer and single SPAD), the use of two detectors with a PBS is important to remove common-mode fluctuations in the emission intensity due to blinking; see Sec.~VI in the Supporting Information for details.
Background signals are also measured from nearby areas of the sample and subtracted from the quantum emitter signals at every $\lambda$/2 angle.

The emission polarization visibility is extracted by fitting the relative polarization intensity signal using a sinusoidal function of the form $A\cos^2(2(\theta-\theta_0)) + B$, where $\theta$ is the rotation angle of the $\lambda$/2 waveplate.
The visibility, $P$, is calculated using eq. (\ref{eq:polarization visibility}), 
where $I_{max}=B+A$ and $I_{min}=B$ are the maximum and minimum intensities.

For comparisons with experiments, the emission polarization visibility distribution for randomly oriented emitters is simulated for both $\pi$ and $\sigma$ transition dipoles.
Using a method based on Ref.~\cite{Fourkas:01}, the polarization visibility associated with each optical transition is simulated by integrating the polarized electric-field intensity for an optical dipole radiation pattern over the objective's collection cone for each polarizer angle.
The simulations further accounted for the random orientations of multiple emitters (we assumed an average of two emitters per spot, based on $\g2$ measurements) and the possibility that the two transitions $D_1$ and $D_2$ associated with a $\sigma$ dipole have different intensities.
See Sec.~VII of the Supporting Information for further details.

\subsubsection{Tools and Instrumentation}

To collect STEM images, 1 mg/mL NC dispersions in hexane are drop-cast onto carbon-coated copper grids. STEM images are taken using a JEOL NEOARM electron microscope with Cold-FEG emission source operated at 200 kV.
STEM images are analyzed using Fiji \cite{Schindelin2019}.
To collect XRD patterns, synthesized NCs are dropcast on test-grade p-Si wafers. The sample is measured using a Rigaku SmartLab diffractometer with Cu K$\alpha$ radiation (40 kV and 44mA) in $\theta$-2$\theta$ geometry. 
Steady-state PL is measured using Princeton Instruments Isoplane spectrometer with a CCD detector. 
The confocal microscope is Nikon Eclipse TE200 which uses Thorlabs motorized stages (MZS500-E Z-Axis Stage and MLS203-1 XY) for coarse adjustments and a fast-steering mirror (1" diameter, protected Al mirror (OIM101 Optics in motion) which uses the analog outputs from the DAQ) for fine adjustments of the lateral position and to generate confocal images.
The objective lens is a Nikon x100 oil immersion objective (Nikon Plan Fluor).
For all measurements, the excitation source is a 405 nm Picoquant LDH-series laser diode. 
The SPADs are Excelitas SPCM-780-14-FC.
The pulsed $\g2$ measurement and lifetime measurements are carried out by routing the detector signals to a PicoHarp 300 TCSPC module.
Time-gating sequences are controlled by a Tektronix AWG520, which triggers the laser source and modulates the control signals applied to a series of rf switches (Mini Circuits ZYSWA-2-50DR).
The time-gated signals are recorded by counter modules in a multipurpose data acquisition system (National Instruments DAQ, PCIe-6323).
For emission polarization measurements, we use a broadband 400-800nm $\lambda$/2 waveplate (AHWP10M-600 Thorlabs) and a Liquid crystal variable retarder with Integrated Controller (Thorlabs LCC2415-VIS) to compensate for the birefringence of the confocal microscope.

\section{Financial interest statement}
The authors declare no competing financial interest.

\begin{acknowledgements}
This work was supported by the National Science Foundation Science and Technology Center for the 
Integration of Modern Optoelectronic Materials on Demand, under Grant No. DMR-2019444.  
Y.E.P. acknowledges support from the Zuckerman STEM Leadership Program and the Israel Council for Higher Education - VATAT. 
S.M.T. acknowledges support from the National Science Foundation Graduate Research Fellowship under Grant No. DGE-1845298.

\end{acknowledgements}

\bibliography{Bib}
\end{document}


\renewcommand{\thefigure}{S\arabic{figure}}


\title{Supporting Information: Room-temperature quantum emission from Cu\textsubscript{Zn}-V\textsubscript{S} defects in  ZnS:Cu colloidal nanocrystals}

\author{Yossef E. Panfil}
\affiliation{Department of Electrical and Systems Engineering, University of Pennsylvania, Philadelphia Pennsylvania 19104, USA}
\author{Sarah M. Thompson}
\altaffiliation{Present address: Sandia National Laboratories, Albuquerque, NM 87185}
\affiliation{Department of Electrical and Systems Engineering, University of Pennsylvania, Philadelphia Pennsylvania 19104, USA}
\author{Gary Chen}
\affiliation{Department of Chemistry, University of Pennsylvania, Philadelphia Pennsylvania 19104, USA}
\author{Jonah Ng}
\affiliation{Department of Electrical and Systems Engineering, University of Pennsylvania, Philadelphia Pennsylvania 19104, USA}
\author{Cherie R. Kagan}
\email[Corresponding authors. ]{lbassett@seas.upenn.edu \& kagan@seas.upenn.edu }
\affiliation{Department of Electrical and Systems Engineering, University of Pennsylvania, Philadelphia Pennsylvania 19104, USA}
\affiliation{Department of Materials Science and Engineering, University of Pennsylvania, Philadelphia Pennsylvania 19104, USA}
\affiliation{Department of Chemistry, University of Pennsylvania, Philadelphia Pennsylvania 19104, USA}
\author{Lee C. Bassett}
\email[Corresponding authors. ]{lbassett@seas.upenn.edu \& kagan@seas.upenn.edu }
\affiliation{Department of Electrical and Systems Engineering, University of Pennsylvania, Philadelphia Pennsylvania 19104, USA}

\date{\today}




\maketitle

\section{XRD Characterization}

X-ray diffraction (XRD) measurements are consistent with NCs having a wurtzite-10H crystal structure.

\begin{figure}[!htb]
\centering
\includegraphics[width=0.5\linewidth]{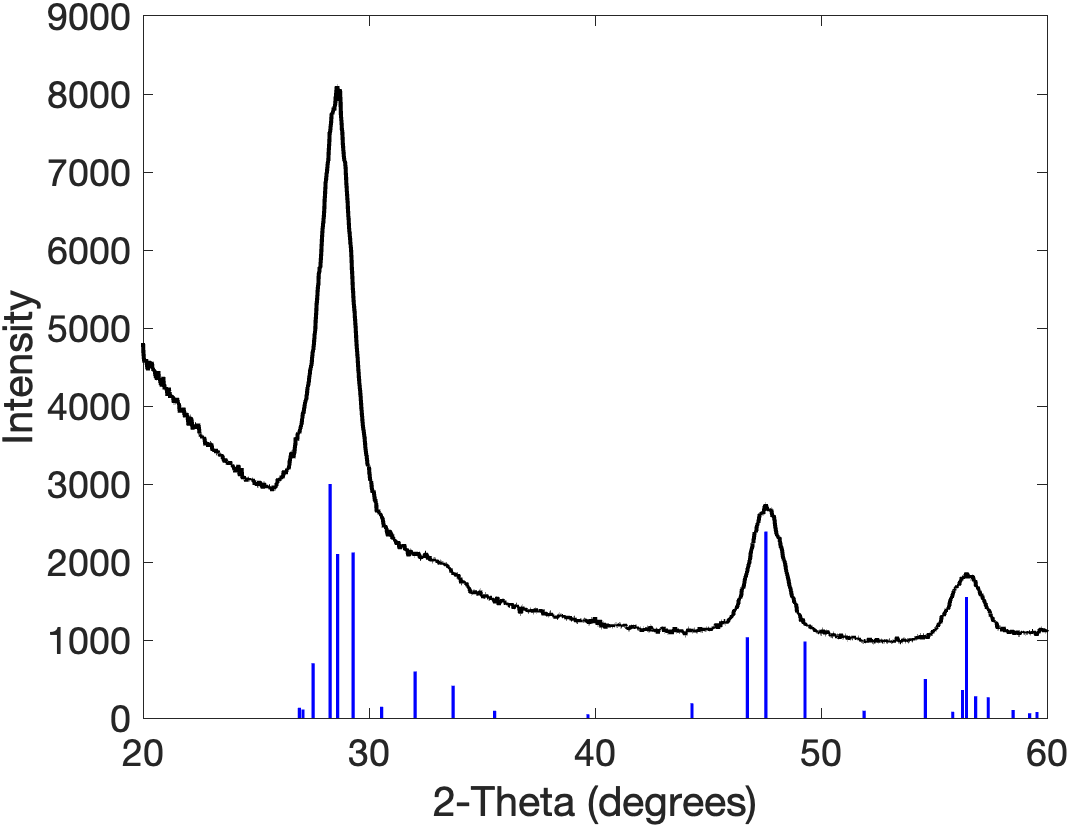}
\caption{
X-ray diffraction (XRD) pattern of the nanocrystals (NCs) showing peaks that correspond to the wurtzite-10H crystal structure, consistent with the observed hexagonal morphology}
\label{fig:dilution}
\end{figure}

\section{Diultion optimization}

Figure~\ref{fig:dilution} shows the late counts image after three dilutions of 1:10 in PMMA and toluene solution (3 wt\%). The intensity and concentration of the spots are reduced after each dilution which proves that the emitting spots are related to the sample and not inherent to the substrate.

\begin{figure}[!htb]
\centering
\includegraphics[width=1\linewidth]{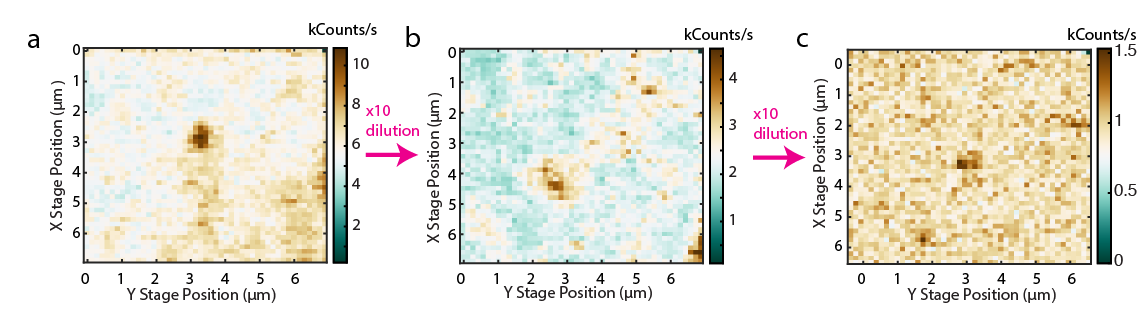}
\caption{Dilution of Z\lowercase{n}S:C\lowercase{u} NC dispersion shown in the late counts image.
(a-c) late counts image after three dilutions of 1:10 in PMMA and toluene solution (3 wt\%). }
\label{fig:dilution}
\end{figure}

\section{Additional quantum-emitter data}

\begin{figure}[!htb]
\centering
\includegraphics[width=1\linewidth]{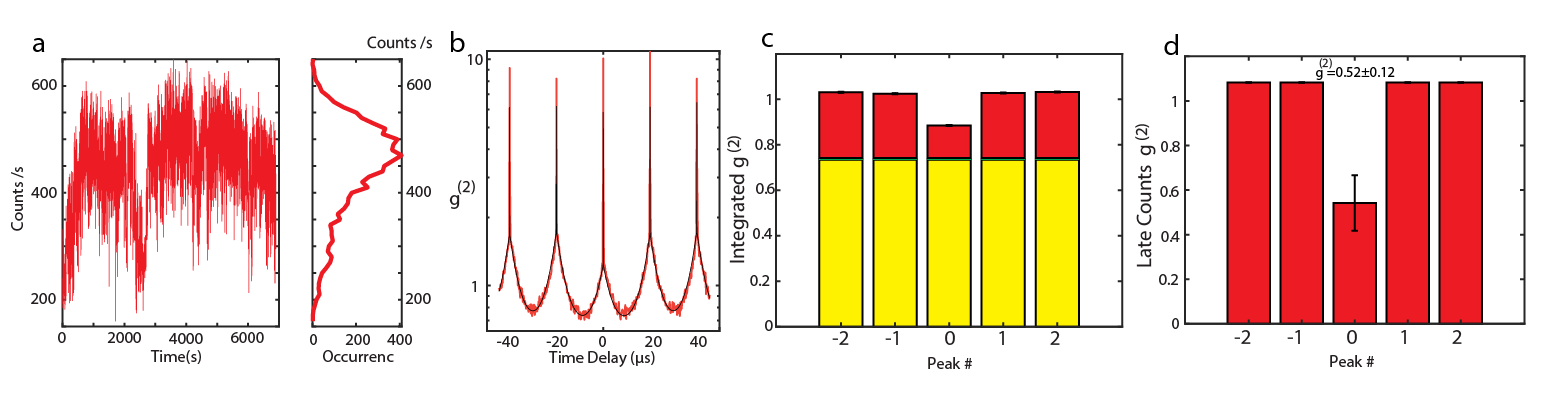}
\caption{Quantum emission from Cu\textsubscript{Zn}-V\textsubscript{S} defects in ZnS:Cu NCs. 
(a) Intensity over 7000 s showing blinking behavior. 
(b) $g^{(2)}$ measurement with peaks separated by the laser repetition rate, showing short and long components with a constant offset (fit in black). 
(c) Integrated counts for short components (green), long components (red), and offset (yellow). 
(d) Corrected $g^{(2)}$ value.}
\label{fig:4t}
\end{figure}

\begin{figure}[!htb]
\centering
\includegraphics[width=1\linewidth]{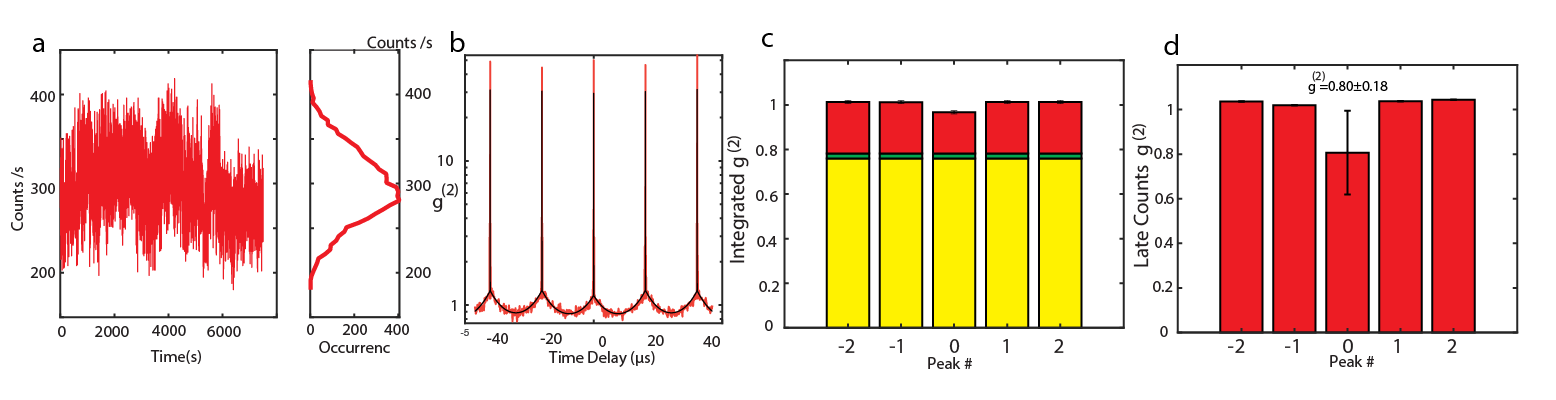}
\caption{Quantum emission from Cu\textsubscript{Zn}-V\textsubscript{S} defects in ZnS:Cu NCs. 
(a) Intensity over 7000 s showing blinking behavior. 
(b) $g^{(2)}$ measurement with peaks separated by the laser repetition rate, showing short and long components with a constant offset (fit in black). 
(c) Integrated counts for short components (green), long components (red), and offset (yellow). 
(d) Corrected $g^{(2)}$ value.}
\label{fig:3c}
\end{figure}

\begin{figure}[!htb]
\centering
\includegraphics[width=1\linewidth]{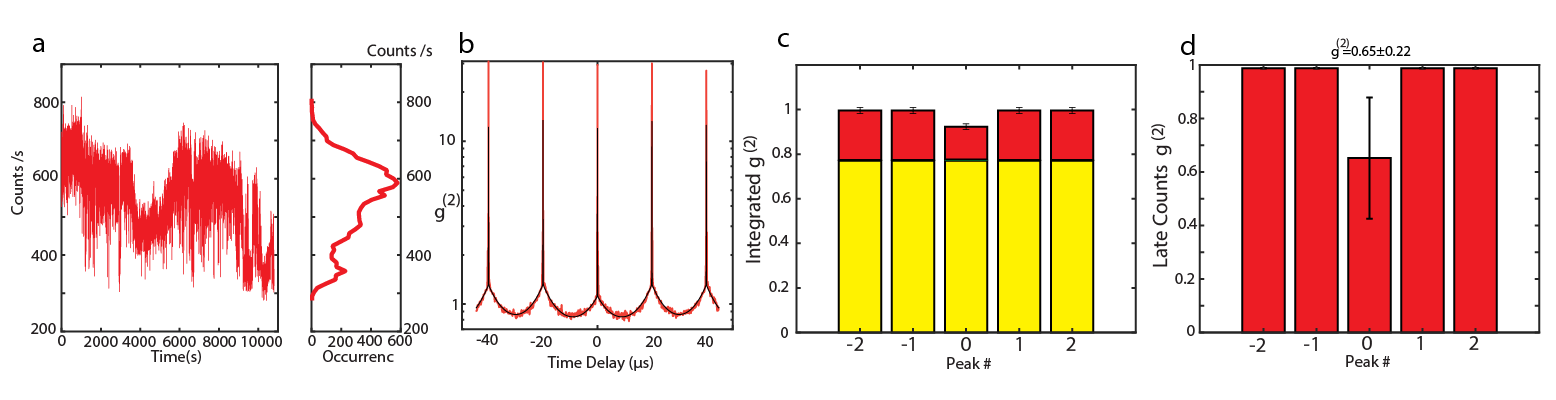}
\caption{Quantum emission from Cu\textsubscript{Zn}-V\textsubscript{S} defects in ZnS:Cu NCs. 
(a) Intensity over 11,000 s showing blinking behavior. 
(b) $g^{(2)}$ measurement with peaks separated by the laser repetition rate, showing short and long components with a constant offset (fit in black). 
(c) Integrated counts for short components (green), long components (red), and offset (yellow). 
(d) Corrected $g^{(2)}$ value.}
\label{fig:5b}
\end{figure}

Figure~\ref{fig:4t}, ~\ref{fig:3c}, and ~\ref{fig:5b} show the $g^{(2)}$ data acquired from other dim spots that appear in the late counts' image. 
Panel (a) shows the intensity variation over 7000-10000 seconds during the $g^{(2)}$ measurement, revealing distinct blinking behavior. Panel (b) presents the $g^{(2)}$ measurement results, with peaks separated by the laser repetition rate, featuring short and long components alongside a constant offset; the data fit is displayed in black.
The $g^{(2)}$ measurements are fitted to the following function:

\begin{align*}
F(\tau) = &\ A_1 e^{-\frac{1}{\tau_1} | \tau - t_0 + 2t_p |} 
+ B_1 e^{-\frac{1}{\tau_2} | \tau - t_0 + 2t_p |} \\
&+ A_2 e^{-\frac{1}{\tau_1} | \tau - t_0 + t_p |} 
+ B_2 e^{-\frac{1}{\tau_2} | \tau - t_0 + t_p |} \\
&+ A_3 e^{-\frac{1}{\tau_1} | \tau - t_0 |} 
+ B_3 e^{-\frac{1}{\tau_2} | \tau - t_0 |} \\
&+ A_4 e^{-\frac{1}{\tau_1} | \tau - t_0 - t_p |} 
+ B_4 e^{-\frac{1}{\tau_2} | \tau - t_0 - t_p |} \\
&+ A_5 e^{-\frac{1}{\tau_1} | \tau - t_0 - 2t_p |} 
+ B_5 e^{-\frac{1}{\tau_2} | \tau - t_0 - 2t_p |} \\
&+ \text{offset}.
\end{align*}

Here, the parameters are interpreted as follows:
\begin{itemize}
    \item $A_1, A_2, A_3, A_4,A_5$ are the amplitudes of the short component.
    \item $B_1, B_2, B_3, B_4, B_5$ are the amplitudes of the long component.
    \item $t_0$ is a shift.
    \item $t_p$ is the inverse of the repetition rate of the laser.
    \item $\tau_1$ and $\tau_2$ are the lifetimes of the short and long components, respectively.
\end{itemize}
Panel (c) highlights the integrated counts under each peak, distinguishing the short component (green), long component (red), and offset/dark counts (yellow). Finally, panel (d) displays the corrected $g^{(2)}$ value.

\clearpage

\section{Autocorrelation corrections}
\section*{Introduction}
The intensity autocorrelation function $g^{(2)}(t)$ is equivalent to the photon detection rate $R$ at time t conditioned on the detection of a photon at t=0, normalized by the time-averaged rate \cite{Brouri:00}.

\begin{equation} \label{g2 correction 1} 
g^{(2)}(\tau) = \frac{R(\tau|0)}{R(\infty)}
\end{equation} 
\\
And for the case of pulsed $g^{(2)}$:
\begin{equation} \label{g2 correction} 
g^{(2)}(j) = \frac{R(j|j=0)}{R(j=\infty)}
\end{equation} 
\\
Where $R(j)$ is the count rate per pulse at pulse $j$.
\\

In case there are several emission sources in the diffraction-limited spot, the rate of detection depends on the source of the first detected photon:
\begin{equation} \label{g2 correction 2}
\begin{split}
R_m(j|j=0) = \frac{S_L}{S_L + S_S + D}(R_L(j|j=0)+S_S+D) + \frac{S_S}{S_L + S_S + D}(R_S(j|j=0)+S_L+D) \\ +\frac{D}{S_L + S_S + D}(S_S+S_L+D),
\end{split}
\end{equation}
where  $R_m(j|j=0)$ is the measured count rate at pulse $j$ conditioned on the detection of a photon at pulse $j=0$. 
$S_L$ is the long component count rate equivalent to $R_L(j=\infty|j=0)$,  $S_S$ is the short component count rate and equivalent to $R_S(j=\infty|j=0)$, and $D$ is the dark count rate which is assumed to be Poissonian and hence independent of detection of a photon in pulse $j=0$.
The prefactor before each parenthesis represents the probability that the first photon originates from the long component emission (in our case the defects), the short component emission (in our case the substrate), or the dark counts.
\\
\\
For $j\neq0$ and assuming that photons collected from adjacent pulses are totally independent, $R_{mL/S}(j\neq0|j=0)=S_{L/S}$, and hence the previous expression will be:
\begin{equation} \label{g2 correction 3}
\begin{split}
R_m(j\neq0|j=0) = \frac{S_L}{S_L + S_S + D}(S_L+S_S+D) + \frac{S_S}{S_L + S_S + D}(S_S+S_L+D) \\ +\frac{D}{S_L + S_S + D}(S_S+S_L+D) 
\end{split}
\end{equation}
and as a consequence,
\begin{equation} \label{g2 correction 4}
\begin{split}
g_m^{(2)}(j\neq0) = \frac{R_m(j\neq0|j=0)}{S_L + S_S + D}=1
\end{split}
\end{equation}
\\
However, for \textit{j=0}:
\begin{equation} \label{g2 correction 5}
\begin{split}
R_m(j=0|j=0) = \frac{S_L}{S_L + S_S + D}(R_L(j=0|j=0)+S_S+D) +\\ \frac{S_S}{S_L + S_S + D}(R_S(j=0|j=0)+S_L+D)+\frac{D}{S_L + S_S + D}(S_S+S_L+D) 
\end{split}
\end{equation}
and as a consequence the $g_m^{(2)}(j=0)$ is:
\begin{equation} \label{g2 correction 6}
\begin{split}
g_m^{(2)}(j=0) = \frac{S_L}{(S_L + S_S + D)^{2}}(R_L(j=0|j=0)+S_S+D) +\\ \frac{S_S}{(S_L + S_S + D)^{2}}(R_S(j=0|j=0)+S_L+D)+\frac{D}{(S_L + S_S + D)^{2}}(S_S+S_L+D) 
\end{split}
\end{equation}

The connection between $R_{L/S}(j=0|j=0)$ to $g_{L/S}^{(2)}(j=0) $ where $R_{L/S}$ and $g^{(2)}_{L/S}$ are the count rate and $g^{(2)}$ in case only long or short component emitters were present in the diffraction-limited spot without dark count as well, correspondingly:
\begin{equation} \label{g2 correction 7}
\begin{split}
R_{L/S}(j=0|j=0)=g_{L/S}^{(2)}(j=0)*S_{L/S} 
\end{split}
\end{equation}
\\
Putting this expression in the previous equation gives:
\begin{equation} \label{g2 correction 8}
\begin{split}
g_m^{(2)}(j=0) = \frac{S_L}{(S_L + S_S + D)^{2}}(g_{L}^{(2)}(j=0)*S_{L}+S_S+D) +\\ \frac{S_S}{(S_L + S_S + D)^{2}}(g_{S}^{(2)}(j=0)*S_{S}+S_L+D)+\frac{D}{(S_L + S_S + D)^{2}}(S_S+S_L+D) 
\end{split}
\end{equation}
Rewriting equation \ref{g2 correction 4} and equation \ref{g2 correction 8} with the color red for terms that contribute to the long component, with blue for terms that contribute to the short component, and orange for terms related to the dark counts that will contribute to the constant offset of the $g^{(2)}$ curve:

\begin{equation} \label{g2 correction 9}
\begin{split}
g_m^{(2)}(j\neq0)  = \frac{S_L}{(S_L + S_S + D)^{2}}({\color{red}S_L}+{\color{blue}S_S}+{\color{orange}D}) + \frac{S_S}{(S_L + S_S + D)^{2}}({\color{blue}S_S}+{\color{red}S_L}+{\color{orange}D}) \\ +\frac{D}{(S_L + S_S + D)^{2}}({\color{orange}S_S+S_L+D}) 
\end{split}
\end{equation}

\begin{equation} \label{g2 correction 10}
\begin{split}
g_m^{(2)}(j=0) = \frac{S_L}{(S_L + S_S + D)^{2}}(g_{L}^{(2)}(j=0)*{\color{red}S_{L}}+{\color{blue}S_S}+{\color{orange}D}) +\\ \frac{S_S}{(S_L + S_S + D)^{2}}(g_{S}^{(2)}(j=0)*{\color{blue}S_{S}}+{\color{red}S_L}+{\color{orange}D})+\frac{D}{(S_L + S_S + D)^{2}}({\color{orange}S_S+S_L+D}) 
\end{split}
\end{equation}

So for the long component the measured $g_{mL}^{(2)}(j=0))$ and $g_{mL}^{(2)}(j\neq0)$ will look like:

\begin{equation} \label{g2 correction 11}
\begin{split}
g_{mL}^{(2)}(j\neq0)  = \frac{S_L}{(S_L + S_S + D)^{2}}(S_L) + \frac{S_S}{(S_L + S_S + D)^{2}}(S_L) 
\end{split}
\end{equation}

\begin{equation} \label{g2 correction 12}
\begin{split}
g_{mL}^{(2)}(j=0) = \frac{S_L}{(S_L + S_S + D)^{2}}(g_{L}^{(2)}(j=0)*S_{L}) + \frac{S_S}{(S_L + S_S + D)^{2}}(S_L) 
\end{split}
\end{equation}

From these expressions it is easy to show that:

\begin{equation} \label{g2 correction 13}
\begin{split}
\frac{g_{mL}^{(2)}(j=0)}{g_{mL}^{(2)}(j\neq0)}  = \frac{S_L*(\frac{g_{L}^{(2)}(j=0)}{g_{L}^{(2)}(j\neq0)})+S_S}{S_L+S_S}
\end{split}
\end{equation}

The expression in \ref{g2 correction 13} is approaching 1 when $S_L<<S_S$. However, when $S_L>>S_S$, the expression in \ref{g2 correction 13} will be $\frac{g_{mL}^{(2)}(j=0)}{g_{mL}^{(2)}(j\neq0)}  \sim \frac{g_{L}^{(2)}(j=0)}{g_{L}^{(2)}(j\neq0)}$.
In our case, since \( S_L \gg S_S \), the correction to the \( g^{(2)} \) data will be small. 

In the following sections, we will detail the methodology used to apply this correction.

\section*{From $g_{mL}^{(2)}$ to the Corrected $g_{L}^{(2)}$}

 The corrected $g_{L}^{(2)}$ value for the central peak is calculated using the following relationship:
\begin{equation}
\begin{split}
g_{L}^{(2)} = \frac{g_{mL}^{(2)} (S_{L}+S_{S}+D)^2 - S_{S}S_{L}}{S_{L}^2}.
\end{split}
\end{equation}

The value of \( g_{mL}^{(2)} \) and its uncertainty is obtained from Figure 2e, as will be explained in the next section. The subsequent section will describe how the parameters \( S_L \), \( S_S \), and \( D \), along with their uncertainties, can be extracted. It is worth mentioning that these parameters can also be independently measured using spatially or time-resolved techniques. Alternatively, they can be estimated from the satellite (\( j \neq 0 \)) peaks of \( g_{mL/S/D}^{(2)} \) in Figure 2e under the assumption that the emission contributing to the satellite peaks is uncorrelated, as detailed in the following sections.

\subsection*{Estimating $g_m^{(2)}$ and Its Uncertainty}

The values of $g_{mL}^{(2)}$, $g_{mS}^{(2)}$, and $g_{mD}^{(2)}$ (both central and side peaks in Figure 2e of the manuscript) and their uncertainties are evaluated from the fitting of five double-exponentials from above. The uncertainty is calculated by propagating the uncertainties of the amplitudes ($A$) and lifetimes ($\tau$) obtained from the fit. Since the integrated peak intensity  is equal to $P = 2 \cdot A \cdot \tau$, the uncertainty in $P$ ($P=g_{mL}^{(2)}$) is determined using the error propagation formula:
\begin{equation}
    \sigma_P = 2 \sqrt{(\tau \cdot \sigma_A)^2 + (A \cdot \sigma_\tau)^2},
\end{equation}
where $\sigma_A$ and $\sigma_\tau$ are the uncertainties in the amplitude and lifetime, respectively, extracted from the fitting process. This formula accounts for the contributions of both parameters to the overall uncertainty in the integrated peak intensity. The uncertainties are reported with 68\% confidence intervals.

\subsection*{Estimating \( S_L \), \( S_S \), and \( D \) and their uncertainties from time-resolved  $g_m^{(2)}$ }
The parameters \(S_L\), \(S_S\), and \(D\) are evaluated from the time-resolved autocorrelation components in Figure 2e of the manuscript as follows: 

Based on eq.10:
\begin{align}
    g_{mL}^{(2)}(j \neq 0) & = \frac{S_L}{(S_L + S_S + D)^2}(S_L) + \frac{S_S}{(S_L + S_S + D)^2}(S_L), \\
    g_{mS}^{(2)}(j \neq 0) & = \frac{S_S}{(S_L + S_S + D)^2}(S_S) + \frac{S_S}{(S_L + S_S + D)^2}(S_L), \\
    g_{mD}^{(2)}(j \neq 0) & = \frac{S_L}{(S_L + S_S + D)^2}(D) + \frac{S_S}{(S_L + S_S + D)^2}(D) + \frac{D}{(S_L + S_S + D)^2}(S_L + S_S + D).
\end{align}

These equations describe how each of the components contributes to the overall time-resolved second-order correlation functions, and each of the three correlation functions is dominated by one of the components (either \( S_L \), \( S_S \), or \( D \)).
These equations can be analytically solved to yield expressions for \( S_L \), \( S_S \), and \( D \) in terms of the time-resolved correlation functions and the total count rate, $S_\mathrm{Total}=S_L+S_S+D$ as follows:

\[
D = \frac{g_{mD}^{(2)} \cdot S_\mathrm{Total}}{1 + \sqrt{g_{mS}^{(2)} + g_{mL}^{(2)}}}
\]

\[
S_L = \frac{S_\mathrm{Total} - D}{1 + \frac{g_{mS}^{(2)}}{g_{mL}^{(2)}}}
\]
\[
S_S = \frac{S_\mathrm{Total} - D}{1 + \frac{g_{mL}^{(2)}}{g_{mS}^{(2)}}}
\]

\subsection*{Uncertainty Propagation for \( S_L \), \( S_S \), and \( D \)}

The uncertainties for \( S_L \), \( S_S \), and \( D \) were estimated using a Monte Carlo approach. Randomized datasets were generated based on the measured $g_{mL}^{(2)}$, $g_{mS}^{(2)}$, $g_{mD}^{(2)}$, and $S_\mathrm{Total}$ values and their uncertainties, and the values of \( S_L \), \( S_S \), and \( D \) were extracted for each iteration. The resulting distributions provided the propagated uncertainties for these parameters.
The value of $S_\mathrm{Total}$ is determined as the time-averaged intensity shown in Figure 2c, with its uncertainty represented by the standard deviation of the intensity distribution.

\subsection*{Averaging and uncertainty calculation of $g_{mL}^{(2)}$ and $g_{mS}^{(2)}$ over 4 side peaks}

When calculating the average values of \( g_{mL}^{(2)} \), \( g_{mS}^{(2)} \), and \( g_{mD}^{(2)} \) from measurements across four peaks, a weighted average is typically used, but when the measured values and uncertainties are similar across the peaks, the arithmetic mean is sufficient. The average of \( g_{mL}^{(2)} \) is given by:
\[
\bar{g}_{mL}^{(2)} = \frac{1}{N} \sum_{i=1}^{N} g_{mL,i}^{(2)},
\]
where \( N = 4 \). The same formula applies for \( \bar{g}_{mS}^{(2)} \):
\[
\bar{g}_{mS}^{(2)} = \frac{1}{N} \sum_{i=1}^{N} g_{mS,i}^{(2)}.
\]
The uncertainty in the mean is calculated as:
\[
\sigma_{\bar{g}_{mL}^{(2)}} = \sqrt{\frac{1}{N} \sum_{i=1}^{N} \sigma_{mL,i}^2},
\]
which simplifies to \( \sigma_{\bar{g}_{mL}^{(2)}} \approx \frac{\sigma_{mL}}{\sqrt{N}} \) if the uncertainties are nearly identical. The same approach applies for \( \sigma_{\bar{g}_{mS}^{(2)}} \).

\subsection*{Error calculation in figure 2f}
Now that we are equipped with the values of \( g_{mL}^{(2)} \), \( S_L \), \( S_S \), and \( D \) along with their uncertainties, we use a Monte Carlo approach to calculate the uncertainty of \( g_{L}^{(2)} \) in Figure 2f. The procedure involves defining normal distributions to model the uncertainties of \( S_L \), \( S_S \), \( D \), and \( g_{mL}^{(2)} \) with their respective means and standard deviations. We then generate \( N \) random samples for each parameter, compute \( g_{L}^{(2)} \) for each set of sampled inputs, and finally analyze the results by determining the mean and standard deviation of the resulting \( g_{L}^{(2)} \) distribution to estimate its central value and uncertainty. The resulting distribution provides the central value and uncertainty of \( g_{L}^{(2)} \).

\section{FWHM and Intensity Distributions of Cu\textsubscript{Zn}-V\textsubscript{S} Spectra}

\begin{figure}[H]
\centering
\includegraphics[width=0.5\textwidth]{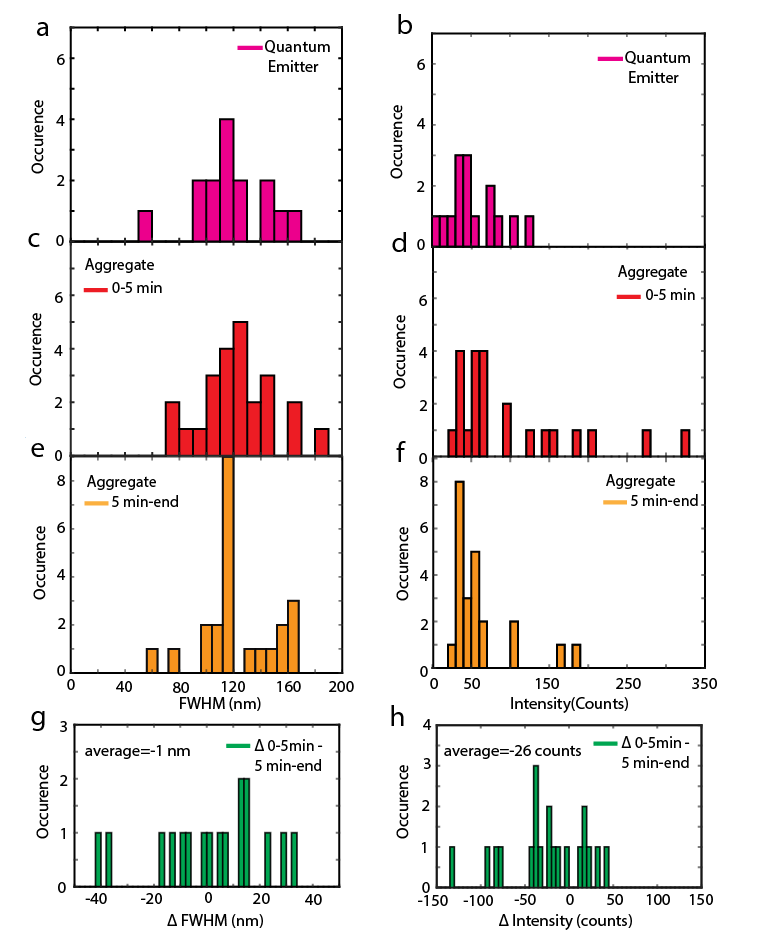}
\caption{\label{fig:PL Spectrums}
\textbf{FWHM and Intensity of Variations under Illumination}
(a) Distribution of the FWHM and intensity (b) in the spectrum of 15 dim spots, some of which are measured to have \textit{g\textsuperscript{(2)}} values less than 1 and others with the same intensity.
(c,d,e,f) Histograms of the FWHM and Intensity for 24 bright spots during the first 5 min (c,d) and after 5 min of illumination (e,f).
(g) Difference in the FWHM and intensity (h) between the first 5-min spectrum and the later spectrum for specific aggregates.
}
\end{figure}

\section{Emission Polarization Visibility of Blinking Quantum Emitters}

The blinking behavior of quantum emitters can introduce intensity fluctuations when only one avalanche photodiode (APD) is used for detection. To eliminate this effect and accurately capture the polarization-dependent characteristics, both linear polarization components are simultaneously measured using a $\lambda$/2 waveplate and a polarization beam splitter (PBS).

Figure \ref{fig:PL Spectrums} shows the intensity of the two polarization components as a function of the $\lambda$/2 waveplate angle. Panel (a) demonstrates the variation in intensity as a function of angle for the two linear polarization components, while panel (b) presents the normalized intensity (by the sum of the two components) as a function of the same angle. This normalization restores the sinusoidal behavior, providing a clearer view of the polarization visibility despite the blinking behavior of the quantum emitters.

\begin{figure}[H]
\centering
\includegraphics[width=0.75\textwidth]{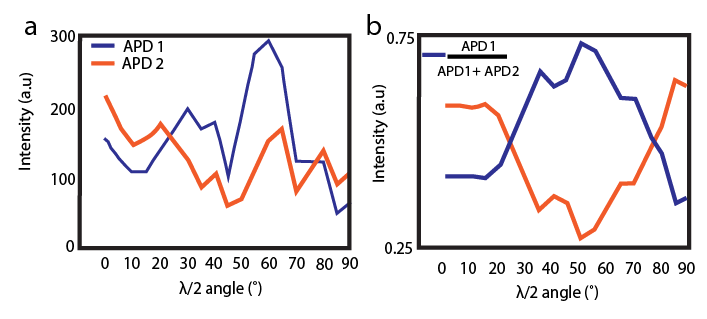}
\caption{\label{fig:PL Spectrums}
\textbf{Emission Polarization Visibility of Blinking Quantum Emitters.} 
\textbf{Emission Polarization Visibility of Blinking Quantum Emitters.} (a) Intensity as a function of the $\lambda$/2 waveplate angle for the two linear polarization components. (b) Intensity normalized by the sum of the two linear polarization components as a function of the $\lambda$/2 waveplate angle.
}
\end{figure}

\section{Emission polarization visibility simulations for randomly oriented emitters} 

We have simulated the expected emission polarization visibility values for randomly oriented Zn:Cu NCs both for the case of $\pi$ and the case of $\sigma$ transition dipoles.
The simulation method is based on \cite{Fourkas:01}.
Knowing the orientation angles of the dipole $\Theta$ and $\Phi$, and the direction of the ray $\theta$ and $\phi$, the electric field components after the objective lens are calculated by matrix multiplication $R\cdot\eta$.
$\eta$ and $R$ are the electric field polarization vector of any given ray and the rotation matrix that makes all fluorescent rays, collected by the objective lens, parallel to the z-axis, respectively. 
Figure \ref{fig:polarization simulation images_1} is demonstrating that:

\begin{figure}[H]
\centering
\includegraphics[width=1\linewidth]{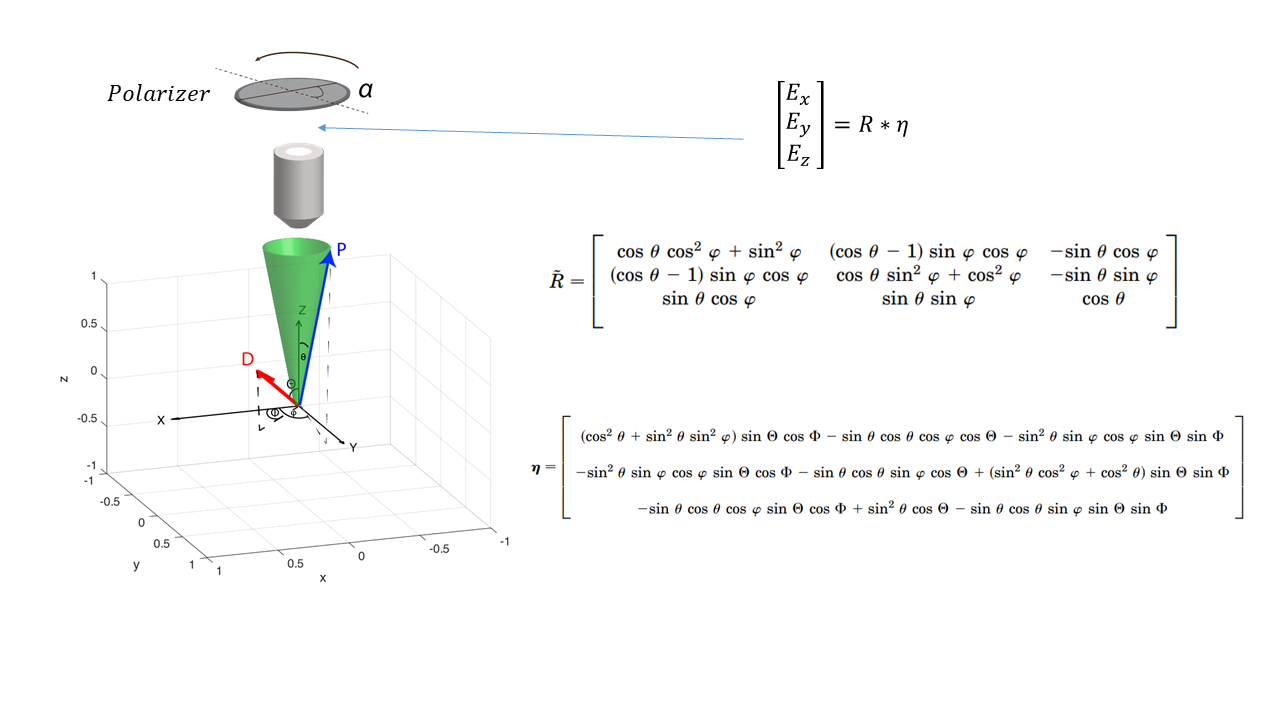}
\caption{Knowing the orientation angles of the dipole $\Theta$ and $\Phi$, and the direction of the ray $\theta$ and $\phi$, the electric field components after the objective lens are calculated by matrix multiplication $R\cdot\eta$.}
\label{fig:polarization simulation images_1}
\end{figure}

The transmitted light through the polarizer at an angle $\alpha$ (the angle between the polarizer and the x-axis) is calculated by projecting the electric field components onto the axis of the polarizer.

\begin{equation} \label{eq polarization 1}
\begin{split}
E_{polarizer} = E_X*cos(\alpha)+E_Y*sin(\alpha)
\end{split}
\end{equation}

The intensity of the emitted light after the objective lens is proportional to the square of the electric field.
As the collection angle of the objective lens is set by the numerical aperture, all the rays within the collection cone are integrated for every polarizer angle $\alpha$.
In our case, we consider an objective with a 1.3 N.A. and a medium with an index of refraction of 1.5, which implies that $\alpha$ = 60$^{\circ}$.
Then, the polarization visibility is calculated by:
\begin{equation} \label{eq polarization 2}
\begin{split}
P=\frac{I_{max}-I_{min}}{I_{max}+I_{min}}
\end{split}
\end{equation}

\begin{figure}[H]
\centering
\includegraphics[width=1\linewidth]{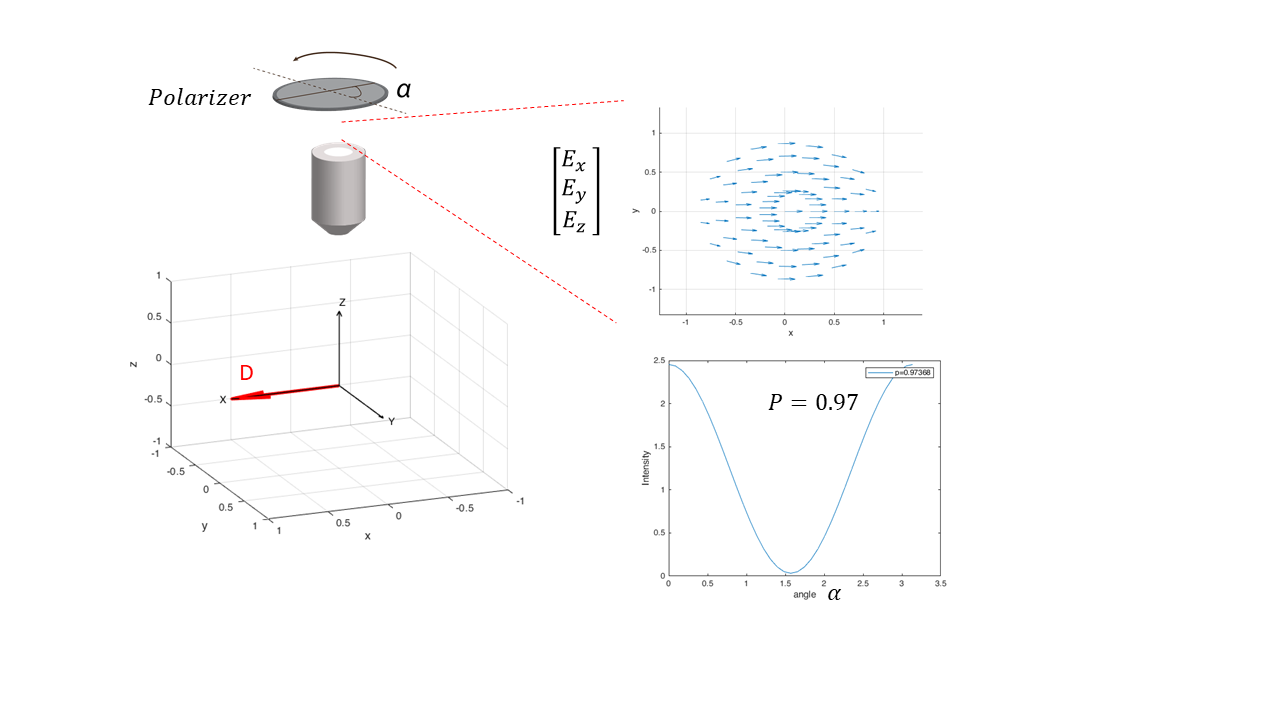}
\caption{The effect of the numerical aperture on the polarization visibility for a dipole in the xy plane.}
\label{fig:polarization simulation images_2}
\end{figure}

\begin{figure}[H]
\centering
\includegraphics[width=1\linewidth]{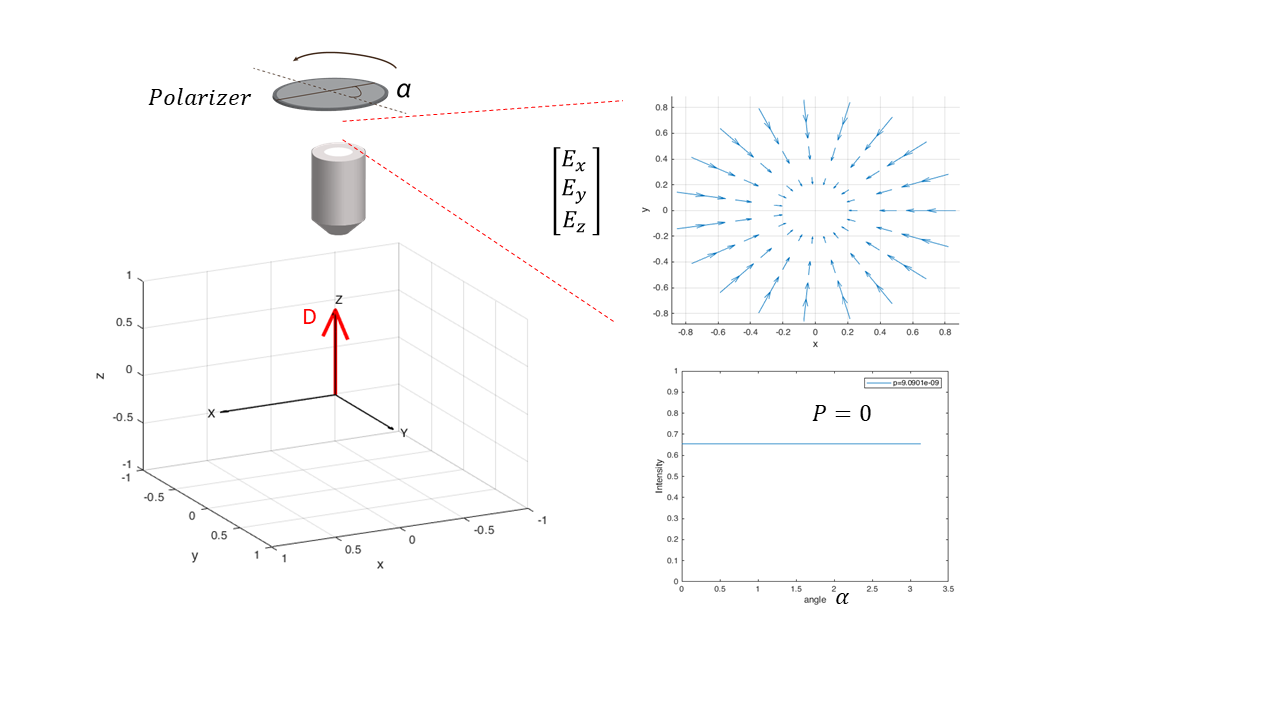}
\caption{The effect of the numerical aperture on the polarization visibility for a dipole along the z-axis.}
\label{fig:polarization simulation images_3}
\end{figure}

\begin{figure}[H]
\centering
\includegraphics[width=1\linewidth]{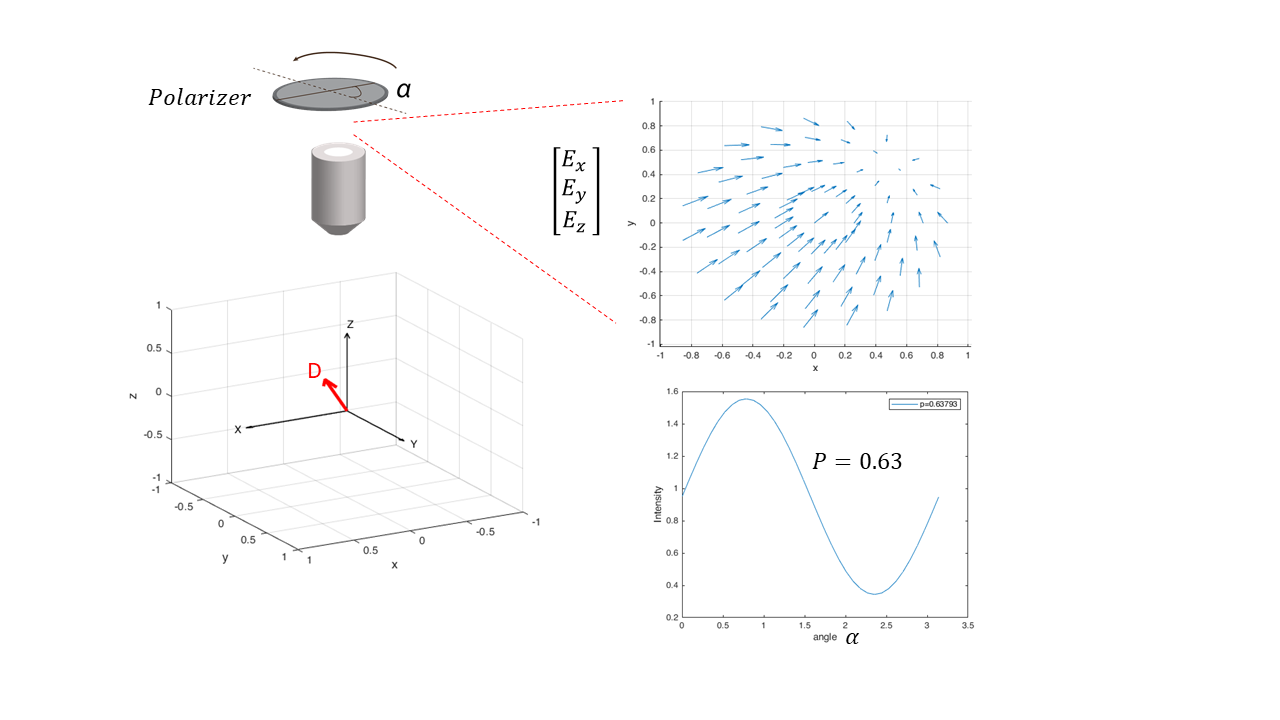}
\caption{The effect of the numerical aperture on the polarization visibility for a dipole at 45$^{\circ}$  from the z-axis.}
\label{fig:polarization simulation images_4}
\end{figure}

The effect of the numerical aperture on the polarization visibility is seen in Figures \ref{fig:polarization simulation images_2}, \ref{fig:polarization simulation images_3}, and \ref{fig:polarization simulation images_4}   where the polarization visibility is calculated for a dipole in the xy plane, a dipole along the z-axis,  and a dipole at 45$^{\circ}$  from the z-axis, correspondingly.
The rays that are not along the optical axis (Z-axis) tend to lower the polarization visibility.

We have sampled spherically uniform distribution of any dipole orientation to simulate the random orientations of the NCs.
Figure \ref{fig:polarization simulation images_5} shows the distribution of the dipole orientations that were sampled to generate the polarization visibility distributions.
In the case of $\sigma$ transition, the two dipoles are perpendicular one to the other, and their contributions to the integrated intensity at every polarizer angle $\alpha$ are added. 

\begin{figure}[H]
\centering
\includegraphics[width=1\linewidth]{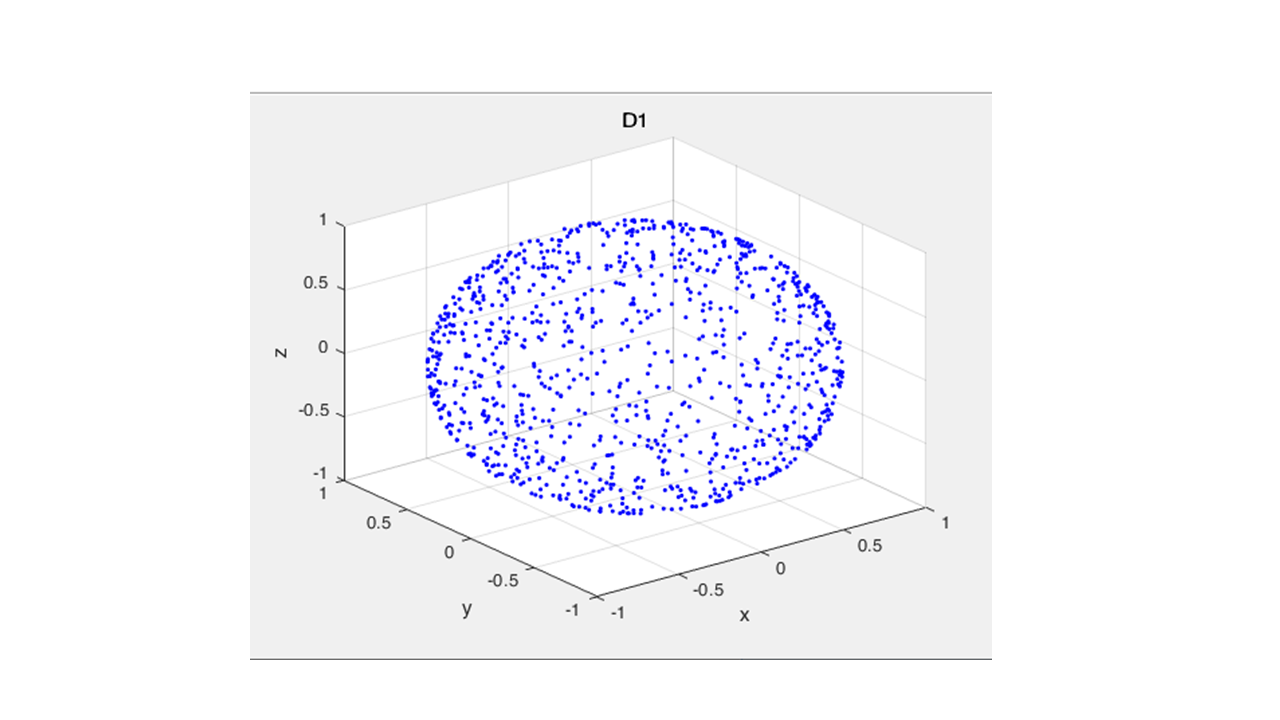}
\caption{The distribution of the dipole orientations that were sampled to generate the polarization visibility distributions}
\label{fig:polarization simulation images_5}
\end{figure}

\section{Emission polarization visibility distributions}

The simulated distribution of emission polarization visibility is presented for both $\pi$ and $\sigma$ dipole transitions in Figure~\ref{fig:polarization simulation images_5}, considering a scenario where one dipole has half the intensity of the other.

\begin{figure}[H]
\centering
\includegraphics[width=0.5\linewidth]{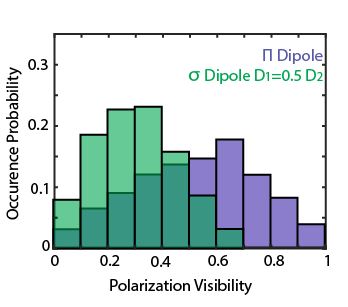}
\caption{The simulated distribution of emission polarization visibility is presented for both $\pi$  and $\sigma$ dipole transitions, considering a scenario where one dipole has half the intensity of the other. This simulation accounts for randomly oriented emitters, representative of the Wurtzite 10-H structure}
\label{fig:polarization simulation images_5}
\end{figure}

\begin{figure}[H]
\centering
\includegraphics[width=1\linewidth]{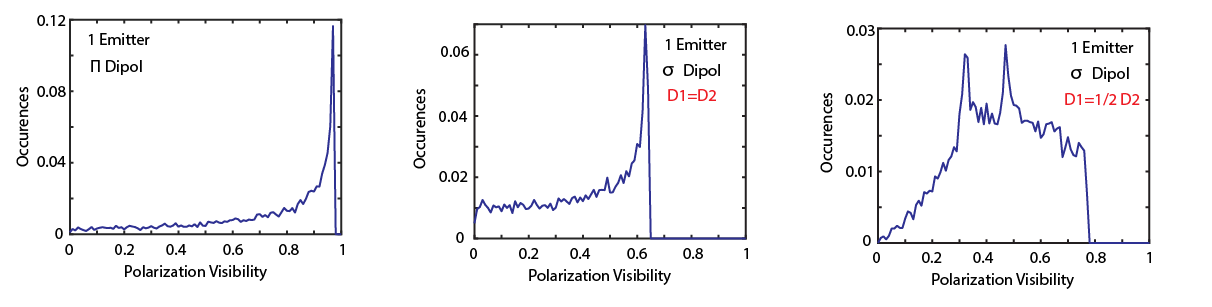}
\caption{The simulated distribution of the emission polarization visibility for a $\pi$ dipole transition, for a $\sigma$ dipole transition, and for a $\sigma$ dipole transition where one of the dipoles has 1/2 of the intensity of the other, for a randomly oriented emitter.}
\label{fig:Figure 4_v5_one_emitter-01}
\end{figure}

In the manuscript, the distributions are shown for the case of two emitters inside the diffraction-limited spot.
Figure \ref{fig:Figure 4_v5_one_emitter-01} illustrates the simulated distribution of emission polarization visibility for a randomly oriented emitter, considering three cases: a $\pi$ dipole transition, a $\sigma$ dipole transition, and a $\sigma$ dipole transition where one of the dipoles has half the intensity of the other.

\clearpage
\bibliography{sample}
